\documentclass[10pt, conference, compsocconf]{IEEEtran}

\usepackage{proof,latexsym, amsmath, amsfonts,amssymb}

\newcommand{\texcomment}[1]{}

\newtheorem{theorem}{\bf Theorem}[section]
\newtheorem{definition}[theorem]{\bf Definition}
\newtheorem{lemma}[theorem]{\bf Lemma}

\newenvironment{reftheorem}[1]{\begin{trivlist}\item[\hskip
      \labelsep{\hspace{1em}\bf Theorem #1:}]\em}{\end{trivlist}}

\newcommand{\aset}[1]{\{{#1}\}}

\newcommand{\sembrack}[1]{[\hspace*{-1.2pt}[#1]\hspace*{-1.2pt}]}

\newcommand{\ttif}[3]{\texttt{if}\;{#1}\;\texttt{then}\;{#2}\;\texttt{else}\;{#3}}

\newcommand{\ttassign}[2]{{#1}\;\texttt{:=}\;{#2}}

\newcommand{\wpre}[2]{\textit{wp}({#1}, {#2})}

\newcommand{\vect}[1]{\overrightarrow{{#1}}}

\begin{document}

\title{Quantitative Information Flow -- Verification Hardness and Possibilities}
 \author{\IEEEauthorblockN{Hirotoshi Yasuoka}
\IEEEauthorblockA{
Graduate School of Information Sciences \\
Tohoku University\\
Sendai, Japan\\
 yasuoka@kb.ecei.tohoku.ac.jp}
\and
\IEEEauthorblockN{Tachio Terauchi}
\IEEEauthorblockA{
Graduate School of Information Sciences \\
Tohoku University\\
Sendai, Japan\\
terauchi@ecei.tohoku.ac.jp}
}

\maketitle

\begin{abstract}

Researchers have proposed formal definitions of quantitative
information flow based on information theoretic notions such as the
Shannon entropy, the min entropy, the guessing entropy, and channel
capacity.  This paper investigates the hardness and possibilities of
precisely checking and inferring quantitative information flow
according to such definitions.

We prove that, even for just comparing two programs on which has the
larger flow, none of the definitions is a k-safety property for any k,
and therefore is not amenable to the self-composition technique that
has been successfully applied to precisely checking non-interference.
We also show a complexity theoretic gap with non-interference by
proving that, for loop-free boolean programs whose non-interference is
coNP-complete, the comparison problem is \#P-hard for all of the
definitions.

For positive results, we show that universally quantifying the
distribution in the comparison problem, that is, comparing two
programs according to the entropy based definitions on which has the
larger flow for all distributions, is a 2-safety problem in general
and is coNP-complete when restricted for loop-free boolean programs.
We prove this by showing that the problem is equivalent to a simple
relation naturally expressing the fact that one program is more
secure than the other.  We prove that the relation also refines the
channel-capacity based definition, and that it can be precisely
checked via the self-composition as well as the ``interleaved''
self-composition technique.
\end{abstract}

\section{Introduction}

\label{sec:introduction}

We consider programs containing high security inputs and low security
outputs.  Informally, the quantitative information flow problem
concerns the amount of information that an attacker can learn about
the high security input by executing the program and observing the low
security output.  The problem is motivated by applications in
information security.  We refer to the classic by
Denning~\cite{denning82} for an overview.

In essence, quantitative information flow measures {\em how} secure, or
insecure, a program is.  Thus, unlike
non-interference~\cite{goguen:sp1982}, that only tells whether a
program is completely secure or not completely secure, a definition of
quantitative information flow must be able to distinguish two programs
that are both interferent but have different degrees of
``secureness.''

For example, consider the following two programs:
\[
\begin{array}{l}
M_1 \equiv \ttif{H = g}{\ttassign{O}{0}}{\ttassign{O}{1}}\\
M_2 \equiv \ttassign{O}{H}
\end{array}
\]
In both programs, $H$ is a high security input and $O$ is a low
security output.  Viewing $H$ as a password, $M_1$ is a prototypical
login program that checks if the guess $g$ matches the
password.\footnote{Here, for simplicity, we assume that $g$ is a
  program constant.  See Section~\ref{sec:prelim} for modeling
  attacker/user (i.e., low security) inputs.}  By executing $M_1$, an
attacker only learns whether $H$ is equal to $g$, whereas she would be
able to learn the entire content of $H$ by executing $M_2$.  Hence, a
reasonable definition of quantitative information flow should assign a
higher quantity to $M_2$ than to $M_1$, whereas non-interference would
merely say that $M_1$ and $M_2$ are both interferent, assuming that
there are more than one possible value of $H$.

Researchers have attempted to formalize the definition of quantitative
information flow by appealing to information theory.  This has
resulted in definitions based on the Shannon
entropy~\cite{denning82,clarkjcs2007,malacaria:popl2007}, the min
entropy~\cite{smith09}, the guessing
entropy~\cite{kopf07,DBLP:conf/sp/BackesKR09}, and channel
capacity~\cite{mccamant:pldi2008,malacaria08,NMS2009}.  Much of the
previous research has focused on information theoretic properties of
the definitions and approximate (i.e., incomplete and/or unsound)
algorithms for checking and inferring quantitative information flow
according to such definitions.

In this paper, we give a verification theoretic and complexity
theoretic analysis of quantitative information flow and investigate
precise methods for checking quantitative information flow.  In
particular, we study the following {\em comparison problem}: Given two
programs $M_1$ and $M_2$, decide if $\mathcal{X}(M_1) \leq
\mathcal{X}(M_2)$.  Here $\mathcal{X}(M)$ denotes the information flow
quantity of the program $M$ according to the quantitative information
flow definition $\mathcal{X}$ where $\mathcal{X}$ is either ${\it
  SE}[\mu]$ (Shannon-entropy based with distribution $\mu$), ${\it
  ME}[\mu]$ (min-entropy based with distribution $\mu$), ${\it
  GE}[\mu]$ (guessing-entropy based with distribution $\mu$), or ${\it
  CC}$ (channel-capacity based).  Note that, obviously, the comparison
problem is no harder than actually computing the quantitative
information flow as we can compare the two numbers once we have
computed $\mathcal{X}(M_1)$ and $\mathcal{X}(M_2)$.

Concretely, we show the following negative results, where
$\mathcal{X}$ is ${\it CC}$, ${\it SE}[\mu]$, ${\it ME}[\mu]$, or ${\it
  GE}[\mu]$ with $\mu$ uniform.
\begin{itemize}
\item Checking if  $\mathcal{X}(M_1) \leq \mathcal{X}(M_2)$ is 
not a $k$-safety property~\cite{terauchi:sas05,DBLP:conf/csfw/ClarksonS08} for any $k$.
\item Restricted to loop-free boolean programs, checking if
  $\mathcal{X}(M_1) \leq \mathcal{X}(M_2)$ is \#P-hard.
\end{itemize}
The results are in stark contrast to non-interference which is known
to be a $2$-safety property in
general~\cite{barthe:csfw04,darvas:spc05} (technically, for the
termination-insensitive case\footnote{We restrict to terminating
  programs in this paper.  (The termination assumption is
  nonrestrictive because we assume safety verification as a blackbox
  routine.)}) and can be shown to be coNP-complete for loop-free
boolean programs (proved in Section~\ref{sec:complex}).  (\#P is known
to be as hard as the entire polynomial hierarchy~\cite{toda91}.)  The
results suggest that precisely inferring (i.e., computing)
quantitative information flow according to these definitions would be
harder than checking non-interference and may require a very different
approach (i.e., not self
composition~\cite{barthe:csfw04,darvas:spc05,terauchi:sas05}).

We also give the following positive results which show checking if the
quantitative information flow of one program is larger than the other
for all distributions according to the entropy-based definitions is
easier.  Below, $\mathcal{Y}$ is ${\it SE}$, ${\it ME}$, or ${\it
GE}$.
\begin{itemize}
\item Checking if $\forall \mu.\mathcal{Y}[\mu](M_1) \leq
  \mathcal{Y}[\mu](M_2)$ is a $2$-safety property.
\item Restricted to loop-free boolean programs, checking if
  $\forall \mu.\mathcal{Y}[\mu](M_1) \leq \mathcal{Y}[\mu](M_2)$ is
  coNP-complete.
\end{itemize}
These results are proven by showing that the problems $\forall
\mu.{\it SE}[\mu](M_1) \leq {\it SE}[\mu](M_2)$, $\forall \mu.{\it
  ME}[\mu](M_1) \leq {\it ME}[\mu](M_2)$, and $\forall \mu.{\it
  GE}[\mu](M_1) \leq {\it GE}[\mu](M_2)$ are all actually equivalent
to a simple $2$-safety relation $R(M_1,M_2)$.  We also show that this
relation refines the channel-capacity based quantitative information
flow, that is, if $R(M_1,M_2)$ then ${\it CC}(M_1) \leq {\it
  CC}(M_2)$.  \texcomment{

The results connecting the different definitions can
be summarized as follows:
\begin{itemize}
\item $\forall \mu.{\it SE}[\mu](M_1) \leq {\it SE}[\mu](M_2)$ if and
  only if $\forall \mu.{\it ME}[\mu](M_1) \leq {\it
    ME}[\mu](M_2)$.\footnote{Technically, this applies only when $M_1$
    and $M_2$ have no low security inputs, because 
    the min-entropy based quantitative information flow is not
    defined for programs with low security inputs.}
\item If $\forall \mu.{\it
  SE}[\mu](M_1) \leq {\it SE}[\mu](M_2)$ then  ${\it CC}(M_1) \leq {\it
  CC}(M_2)$.
\end{itemize}
}

The fact that $R(M_1,M_2)$ is a $2$-safety property implies that it
can be reduced to a safety problem via self composition.  This leads
to a new approach to precisely checking quantitative information flow
that leverages recent advances in automated software
verification~\cite{DBLP:conf/popl/BallR02,DBLP:conf/popl/HenzingerJMS02,mcmillan:cav06,DBLP:journals/sttt/BeyerHJM07}.
Briefly, given $M_1$ and $M_2$, $R(M_1,M_2)$ means that $M_1$ is at
least as secure as $M_2$ for all distributions while $\neg R(M_1,M_2)$
means that there must be a distribution in which $M_1$ is less secure
than $M_2$, according to the entropy-based definitions of quantitative
information flow.  Therefore, by deciding $R(M_1,M_2)$, we can measure
the security of the program $M_1$ relative to another {\em
  specification} program $M_2$.  Note that this is useful even when
$M_1$ and $M_2$ are ``incomparable'' by $R$, that is, when $\neg
R(M_1, M_2)$ and $\neg R(M_2, M_1)$.  See Section~\ref{sec:qbysc} for
the details.

The rest of the paper is organized as follows.
Section~\ref{sec:prelim} reviews the existing information-theoretic
definitions of quantitative information flow.
Section~\ref{sec:qifnksafe} proves the hardness of their comparison
problems and thus shows the hardness of precisely inferring
quantitative information flow according to these definitions.
Section~\ref{sec:qshan} introduces the relation $R$ and proves it
equivalent to the comparison problems for the entropy-based
definitions with their distributions universally quantified.  The
section also shows that this is a $2$-safety property and is easier to
decide than the non-universally-quantified comparison problems, and
suggests a self-composition based method for precisely checking
quantitative information flow.  Section~\ref{sec:related} discusses
related work, and Section~\ref{sec:concl} concludes.
Appendix~\ref{sec:lemdefs} contains the supporting lemmas and
definitions for the proofs appearing in the main text.  The omitted
proofs appear in Appendix~\ref{sec:proofs}.

\section{Preliminaries}

\label{sec:prelim}

We introduce the information theoretic definitions of quantitative
information flow that have been proposed in literature.  First, we
review the notion of the {\em Shannon entropy}~\cite{shannon48},
$\mathcal{H}[\mu](X)$, which is the average of the information
content, and intuitively, denotes the uncertainty of the random
variable $X$.
\begin{definition}[Shannon Entropy]
  Let $X$ be a random variable with sample space $\mathbb X$ and $\mu$
  be a probability distribution associated with $X$ (we write $\mu$
  explicitly for clarity).  The Shannon entropy of $X$ is defined as
\[
\mathcal{H}[\mu](X)=\sum_{x\in\mathbb{X}} \mu(X=x)\log\frac{1}{\mu(X=x)}
\]
(The logarithm is in base 2.)
\end{definition}
Next, we define {\em conditional entropy}.  Informally, the conditional
entropy of $X$ given $Y$ denotes the uncertainty of $X$ after knowing
$Y$.
\begin{definition}[Conditional Entropy]
Let $X$ and $Y$ be random variables with sample spaces $\mathbb X$ and
$\mathbb Y$, respectively, and $\mu$ be a probability distribution
associated with $X$ and $Y$.  Then, the conditional entropy of $X$
given $Y$, written $\mathcal{H}[\mu](X|Y)$ is defined as
\[
\mathcal{H}[\mu](X|Y)=\sum_{y\in\mathbb Y} \mu(Y=y) \mathcal{H}[\mu](X|Y=y)
\]
where
\[
\begin{array}{l}
\mathcal{H}[\mu](X|Y=y) \\
\hspace{2em}=\sum_{x\in\mathbb X} \mu(X=x|Y=y)\log\frac{1}{\mu(X=x|Y=y)} \\
\mu(X=x|Y=y)=\frac{\mu(X=x,Y=y)}{\mu(Y=y)}
\end{array}
\]
\end{definition}
Next, we define (conditional) mutual information.  Intuitively, the
conditional mutual information of $X$ and $Y$ given $Z$ represents the
mutual dependence of $X$ and $Y$ after knowing $Z$.
\begin{definition}[Mutual Information]
Let $X, Y$ and $Z$ be random variables and $\mu$ be an associated
probability distribution.\footnote{We abbreviate sample spaces of
  random variables when they are clear from the context.}  Then, the
conditional mutual information of $X$ and $Y$ given $Z$ is
defined as
\[
\begin{array}{rcl}
\mathcal{I}[\mu](X;Y|Z)&=&\mathcal{H}[\mu](X|Z)-\mathcal{H}[\mu](X|Y,Z)\\
&=&\mathcal{H}[\mu](Y|Z)-\mathcal{H}[\mu](Y|X,Z)
\end{array}
\]
\end{definition}

Let $M$ be a program that takes a high security input $H$ and a low
security input $L$, and gives the low security output $O$.  For
simplicity, we restrict to programs with just one variable of each
kind, but it is trivial to extend the formalism to multiple variables
(e.g., by letting the variables range over tuples).  Also, for the
purpose of the paper, unobservable (i.e., high security) outputs are
irrelevant, and so we assume that the only program output is the low
security output.  Let $\mu$ be a probability distribution over the
values of $H$ and $L$.  Then, the semantics of $M$ can be defined by
the following probability equation.  (We restrict to terminating
deterministic programs in this paper.)
\[
\mu(O = o) = \sum_{\scriptsize \begin{array}{l}h,\ell \in \mathbb{H}, \mathbb{L}\\ M(h,\ell) = o\end{array}} \mu(H = h, L = \ell)
\]
Note that we write $M(h,\ell)$ to denote the low security output of
the program $M$ given inputs $h$ and $\ell$.  Now, we are ready to
introduce the Shannon-entropy based definition of quantitative
information flow (QIF)~\cite{denning82,clarkjcs2007,malacaria:popl2007}.
\begin{definition}[Shannon-Entropy-based QIF]
\label{def:se}
Let $M$ be a program with high security input $H$, low security input
$L$, and low security output $O$.  Let $\mu$ be a distribution over
$H$ and $L$.  Then, the Shannon-entropy-based quantitative information
flow is defined
\[
\begin{array}{rcl}
{\it SE}[\mu](M) & = & \mathcal{I}[\mu](O;H|L) \\
 & = & \mathcal{H}[\mu](H|L)-\mathcal{H}[\mu](H|O,L)
\end{array}
\]
\end{definition}
Intuitively, $\mathcal{H}[\mu](H|L)$ denotes the initial uncertainty
knowing the low security input and $\mathcal{H}[\mu](H|O,L)$ denotes
the remaining uncertainty after knowing the low security output.

As an example, consider the programs $M_1$ and $M_2$ from
Section~\ref{sec:introduction}.  For concreteness, assume that $g$ is
the value $01$ and $H$ ranges over the space $\aset{00,01,10,11}$.
Let $U$ be the uniform distribution over $\aset{00,01,10,11}$, that
is, $U(h) = 1/4$ for all $h \in \aset{00,01,10,11}$.  The results are
as follows.
\[
\begin{array}{rl}
{\it SE}[U](M_1)&=\mathcal{H}[U](H)-\mathcal{H}[U](H|O)\\
&=\log 4-\frac{3}{4}\log{3}\\
&\approx .81128\\&\\

{\it SE}[U](M_2)&=\mathcal{H}[U](H)-\mathcal{H}[U](H|O)\\
&=\log 4-\log 1\\
&=2
\end{array}
\]
Consequently, we have that ${\it SE}[U](M_1) \leq {\it SE}[U](M_2)$,
but ${\it SE}[U](M_2) \not\leq {\it SE}[U](M_1)$.  That is, $M_1$ is
more secure than $M_2$ (according to the Shannon-entropy based
definition with uniformly distributed inputs), which agrees with our
intuition.

Let us recall the notion of
non-interference~\cite{DBLP:conf/sosp/Cohen77,goguen:sp1982}.
\begin{definition}[Non-intereference]
A program $M$ is said to be non-interferent iff for any $h,h'\in
\mathbb{H}$ and $\ell\in\mathbb{L}$, $M(h,\ell)=M(h',\ell)$.
\end{definition}

It is worth noting that non-interference can be formalized as a
special case of the Shannon-entropy based quantitative information
flow where the flow quantity is zero.
\begin{theorem}\label{thm:ReqNI}
  Let $M$ be a program that takes high-security input $H$,
  low-security input $L$, and returns low-security output $O$.  Then,
  $M$ is non-interferent if and only if $\forall \mu.{\it
    SE}[\mu](M)=0$.
\end{theorem}
The above theorem is complementary to the one proven by Clark et
al.~\cite{clark05} which states that for any $\mu$ such that $\mu(H =
h,L = \ell) > 0$ for all $h \in \mathbb{H}$ and $\ell \in \mathbb{L}$,
${\it SE}[\mu](M)=0$ iff $M$ is non-interferent.

Next, we introduce the {\em min entropy}, which Smith~\cite{smith09}
recently suggested as an alternative measure for quantitative information
flow.
\begin{definition}[Min Entropy]
Let $X$ and $Y$ be random variables, and $\mu$ be an associated probability
distribution.  Then, the min entropy of $X$ is defined
\[
\mathcal{H}_\infty[\mu](X)=\log\frac{1}{\mathcal{V}[\mu](X)}
\]
and the conditional min entropy of $X$ given $Y$ is defined
\[
\mathcal{H}_\infty[\mu](X|Y)=\log\frac{1}{\mathcal{V}[\mu](X|Y)}
\]
where
\[
\begin{array}{rl}
\mathcal{V}[\mu](X)&=\max_{x\in\mathbb X} \mu(X=x)\\
\mathcal{V}[\mu](X|Y=y)&=\max_{x\in\mathbb X} \mu(X=x|Y=y)\\
\mathcal{V}[\mu](X|Y)&=\sum_{y\in\mathbb Y} \mu(Y=y) \mathcal{V}[\mu](X|Y=y)
\end{array}
\]
\end{definition}

Intuitively, $\mathcal{V}[\mu](X)$ represents the highest probability
that an attacker guesses $X$ in a single try.  We now define the
min-entropy-based definition of quantitative information flow.

\begin{definition}[Min-Entropy-based QIF]
\label{def:me}
Let $M$ be a program with high security input $H$, low security input
$L$, and low security output $O$.  Let $\mu$ be a distribution over
$H$ and $L$.  Then, the min-entropy-based quantitative information
flow is defined
\[
{\it ME}[\mu](M)=\mathcal{H}_\infty[\mu](H|L)-\mathcal{H}_\infty[\mu](H|O,L)
\]
\end{definition}

Whereas Smith~\cite{smith09} focused on programs lacking low security
inputs, we extend the definition to programs with low security inputs
in the definition above.  It is easy to see that our definition
coincides with Smith's for programs without low security inputs.
Also, the extension is arguably natural in the sense that we simply
take the conditional entropy with respect to the distribution over the
low security inputs.

Computing the min-entropy based quantitative information flow for our
running example programs $M_1$ and $M_2$ from
Section~\ref{sec:introduction} with the uniform distribution, we
obtain,
\[
\begin{array}{rl}
{\it ME}[U](M_1)&=\mathcal{H}_\infty[U](H)-\mathcal{H}_\infty[U](H|O)\\
&=\log 4-\log 2\\
&=1\\&\\
{\it ME}[U](M_2)&=\mathcal{H}_\infty[U](H)-\mathcal{H}_\infty[U](H|O)\\
&=\log 4 -\log 1\\
&=2
\end{array}
\]
Again, we have that ${\it ME}[U](M_1) \leq {\it ME}[U](M_2)$ and ${\it
  ME}[U](M_2) \not\leq {\it ME}[U](M_1)$, and so $M_2$ is deemed less
secure than $M_1$.

The third definition of quantitative information flow treated in this
paper is the one based on the guessing entropy~\cite{Massey94}, that
is also recently proposed in
literature~\cite{kopf07,DBLP:conf/sp/BackesKR09}.
\begin{definition}[Guessing Entropy]
Let $X$ and $Y$ be random variables, and $\mu$ be an associated probability
distribution.  Then, the guessing entropy of $X$ is defined
\[
\mathcal{G}[\mu](X)=\sum_{1\le i\le m}i\times\mu(X=x_i)
\]
where $\aset{x_1,x_2,\dots,x_{m}} = \mathbb{X}$ and
$\forall i,j.i\le j\Rightarrow \mu(X=x_i)\ge\mu(X=x_j)$.

The conditional guessing entropy of $X$ given $Y$ is defined
\[
\mathcal{G}[\mu](X|Y)=\sum_{y\in{\mathbb Y}}\mu(Y=y)\sum_{1\le i\le m}i\times\mu(X=x_i|Y=y)
\]
where $\aset{x_1,x_2,\dots,x_{m}} = \mathbb{X}$ and $\forall i,j.i\le
j\Rightarrow \mu(X=x_i|Y=y)\ge\mu(X=x_j|Y=y)$.
\end{definition}

Intuitively, $\mathcal{G}[\mu](X)$ represents the average number of
times required for the attacker to guess the value of $X$.  We now
define the guessing-entropy-based quantitative information flow.

\begin{definition}[Guessing-Entropy-based QIF]
\label{def:ge}
Let $M$ be a program with high security input $H$, low security input
$L$, and low security output $O$.  Let $\mu$ be a distribution over
$H$ and $L$.  Then, the guessing-entropy-based quantitative
information flow is defined
\[
{\it GE}[\mu](M)=\mathcal{G}[\mu](H|L)-\mathcal{G}[\mu](H|O,L)
\]
\end{definition}

Like with the min-entropy-based definition, the previous research on
guessing-entropy-based quantitative information flow only considered
programs without low security
inputs~\cite{kopf07,DBLP:conf/sp/BackesKR09}.  But, it is easy to see
that our definition with low security inputs coincides with the
previous definitions for programs without low security inputs.  Also,
as with the extension for the min-entropy-based definition, it simply
takes the conditional entropy over the low security inputs.

We test {\it GE} on the running example from
Section~\ref{sec:introduction} by calculating the quantities for the
programs $M_1$ and $M_2$ with the uniform distribution.  
\[
\begin{array}{rl}
{\it GE}[U](M_1) & =\mathcal{G}[U](H)-\mathcal{G}[U](H|O)\\
&=\frac{5}{2} - \frac{7}{4}\\
& = 0.75\\\\
{\it GE}[U](M_2)& = \mathcal{G}[U](H)-\mathcal{G}[U](H|O)\\
&=\frac{5}{2} - 1\\
& = 1.5
\end{array}
\]
Therefore, we again have that ${\it GE}[U](M_1) \leq {\it GE}[U](M_2)$
and ${\it GE}[U](M_2) \not\leq {\it GE}[U](M_1)$, and so $M_2$ is
considered less secure than $M_1$, even with the guessing-entropy
based definition with the uniform distribution.

The fourth and the final existing definition of quantitative
information flow that we introduce in this paper is the one based on
{\em channel capacity}~\cite{mccamant:pldi2008,malacaria08,NMS2009},
which is simply defined to be the maximum of the Shannon-entropy based
quantitative information flow over the distribution.
\begin{definition}[Channel-Capacity-based QIF]

Let $M$ be a program with high security input $H$, low security input
$L$, and low security output $O$.  Then, the channel-capacity-based
quantitative information flow is defined
\[
{\it CC}(M)=\max_\mu\mathcal{I}[\mu](O;H|L)
\]
\end{definition}

Unlike the Shannon-entropy based, the min-entropy based, and the
guessing-entropy based definitions, the channel-capacity based
definition of quantitative information flow is not parameterized by a
distribution over the inputs.  As with the other definitions, let us
test the definition on the running example from
Section~\ref{sec:introduction} by calculating the quantities for the
programs $M_1$ and $M_2$:
\[
\begin{array}{rl}
{\it CC}(M_1)&=\max_\mu\mathcal{I}[\mu](O;H)\\
&=1\\&\\
{\it CC}(M_2)&=\max_\mu\mathcal{I}[\mu](O;H)\\
&=2
\end{array}
\]
As with the entropy-based definitions (with the uniform distribution),
we have that ${\it CC}(M_1) \leq {\it CC}(M_2)$ and ${\it CC}(M_2)
\not\leq {\it CC}(M_1)$, that is, the channel-capacity based
quantitative information flow also says that $M_2$ is less secure than
$M_1$.

\texcomment{
In the absence of low security inputs, 
the probability distribution $\mu$ that maximizes the Shannon-entropy
based information flow is exactly the one that satisfies
the following condition:
\[
\forall o\in\mathbb{O}.\mu(O=o)=\frac{1}{|\mathbb{O}|}
\] 
where $\mathbb{O}$ is a set of all outputs.  This result is noted
in \cite{smith09,malacaria08}.  We note that the min-entropy-based
flow of $M_1$ and $M_2$ and the channel-capacity-based flow of $M_1$
and $M_2$ are same, respectively.  This result is noted in
\cite{smith09}.  Typically, for low security input free programs, the
min-entropy-based flow with uniformly distributed input is the same as
channel-capacity-based flow.
}

\section{Hardness of Comparison Problems}

\label{sec:qifnksafe}

We investigate the hardness of deciding the following {\em comparison
  problem} $C_{\it SE}[\mu]$: Given programs $M_1$ and $M_2$ having
the same input domain, decide if ${\it SE}[\mu](M_1) \leq {\it
  SE}[\mu](M_2)$.  Because we are interested in hardness, we focus on
the case where $\mu$ is the uniform distribution $U$.  That is, the
results we prove for the specific case applies to the general case.
Also note that the comparison problem is no harder than actually
computing the quantitative information flow because we can compare
${\it SE}[\mu](M_1)$ and ${\it SE}[\mu](M_2)$ if we know their actual
values.

Likewise, we study the hardness of the comparison problem $C_{\it
  ME}[\mu]$, defined to be the problem ${\it ME}[\mu](M_1) \leq {\it
  ME}[\mu](M_2)$, $C_{\it GE}[\mu]$, defined to be the problem ${\it
  GE}[\mu](M_1) \leq {\it GE}[\mu](M_2)$, and $C_{\it CC}$, defined to
be the problem ${\it CC}(M_1) \leq {\it CC}(M_2)$.  As with $C_{\it
  SE}[\mu]$, we require the two programs to share the same input
domain for these problems.

We show that none of these comparison problems are $k$-safety problems
for any $k$.  Informally, a program property is said to be a {\em
  $k$-safety}
property~\cite{terauchi:sas05,DBLP:conf/csfw/ClarksonS08} if it can be
refuted by observing $k$ number of (finite) execution traces.  A
$k$-safety problem is the problem of checking a $k$-safety property.
Note that the standard safety property is a $1$-safety property.  An
important property of a $k$-safety problem is that it can be reduced
to a standard safety (i.e., $1$-safety) problem, such as the
unreachability problem, via a simple program transformation called {\em
  self composition}~\cite{barthe:csfw04,darvas:spc05}.

It is well-known that non-interference is a $2$-safety
property,\footnote{It is also well known that it is not a $1$-safety
  property~\cite{mclean:sp94}.} and this has enabled its precise
checking via a reduction to a safety problem via self composition and
piggybacking on advances in automated safety verification
methods~\cite{terauchi:sas05,naumann:esorics06,unno:plas2006}.
Unfortunately, the results in this section imply that quantitative
information flow inference problem is unlikely to receive the same
benefits.

Because we are concerned with properties about pairs of programs
(i.e., comparison problems), we extend the notion of $k$-safety to
properties refutable by observing $k$ traces from each of the two
programs.  More formally, we say that the comparison problem $C$ is a
$k$-safety property if $(M_1, M_2) \not\in C$ implies that there
exists $T_1 \subseteq \sembrack{M_1}$ and $T_2 \subseteq
\sembrack{M_2}$ such that
\begin{itemize}
\item[(1)] $|T_1| \leq k$
\item[(2)] $|T_2| \leq k$
\item[(3)] $\forall M_1',M_2'.T_1 \subseteq \sembrack{M_1'} \wedge
T_2 \subseteq \sembrack{M_2'} \Rightarrow 
(M_1', M_2') \not\in C$
\end{itemize}
In the above, $\sembrack{M}$ denotes the semantics (i.e., traces)
of $M$, represented by the set of input/output pairs
$\aset{((h,\ell),o) \mid h \in \mathbb{H}, \ell \in \mathbb{L}, o =
  M(h,\ell)}$.

We now state the main results of the section.  (Recall that $U$
denotes the uniform distribution.)   We sketch the main idea of the
proofs.  All proofs are by contradiction.  Let $C$ be the comparison
problem in the statement and suppose $C$ is $k$-safety.  Let $(M_1, M_2)
\not\in C$.  Then, we have $T_1 \subseteq \sembrack{M_1}$ and $T_2
\subseteq \sembrack{M_2}$ satisfying the properties (1), (2), and (3)
above.  From this, we construct $\bar{M_1}$ and $\bar{M_2}$ such that
$T_1 \subseteq \sembrack{\bar{M_1}}$ and $T_2 \subseteq
\sembrack{\bar{M_2}}$ and $(\bar{M_1}, \bar{M_2}) \in C$ to obtain
the contradiction.
\begin{theorem}
$C_{\it SE}[U]$ is not a $k$-safety property for any $k > 0$.
\label{thm:seks}
\end{theorem}

\begin{theorem}
$C_{\it ME}[U]$ is not a $k$-safety property for any $k > 0$.
\label{thm:meks}
\end{theorem}

\begin{theorem}
$C_{\it GE}[U]$ is not a $k$-safety property for any $k>0$.
\label{thm:geks}
\end{theorem}

\begin{theorem}
$C_{\it CC}$ is not a $k$-safety property for any $k > 0$.
\label{thm:ccks}
\end{theorem}

\subsection{Bounding the Domains}

The notion of $k$-safety property, like the notion of safety property
from where it extends, is defined over all programs regardless of
their size.  (For example, non-interference is a $2$-safety property
for all programs and unreachability is a safety property for all
programs.)  But, it is easy to show that the comparison problems would
become ``$k$-safety'' properties if we constrained and bounded the
input domains because then the size of the semantics (i.e., the
input/output pairs) of such programs would be bounded by
$|\mathbb{H}|\times|\mathbb{L}|$.  In this case, the problems are at
most $|\mathbb{H}|\times|\mathbb{L}|$-safety.\footnote{It is possible
to get a tighter bound for the channel-capacity based definition by
also bounding the size of the output domain.}  However, these bounds
are high for all but very small domains, and are unlikely to lead to a
practical verification method.

\subsection{Proof of Theorem~\ref{thm:seks}}

We discuss the details of the proof of Theorem~\ref{thm:seks}.  The
proofs of Theorems~\ref{thm:meks}, \ref{thm:geks}, \ref{thm:ccks} are
deferred to Appendix~\ref{sec:proofs}.  

For contradiction, suppose $C_{\it SE}[U]$ is a $k$-safety property.
Let $M$ and $M'$ be programs having the same input domain such that
$(M,M')\not\in C_{\it SE}[U]$.  Then, it must be the case that there
exist $T \subseteq \sembrack{M}$ and $T' \subseteq \sembrack{M'}$ such
that $|T| \leq k$, $|T'| \leq k$, and $\forall M_c,M_c'.T \subseteq
\sembrack{M_c}\wedge T' \subseteq \sembrack{M_c'} \Rightarrow (M_c, M_c')
\not\in C_{\it SE}[U]$.  

Let
\[
\begin{array}{l}
T=\aset{(h_1,o_1),(h_2,o_2),\dots,(h_i,o_i)}\\
T'=\aset{(h_1',o_1'),(h_2',o_2'),\dots,(h_j',o_j')}
\end{array}
\]
where $i,j\le k$.  Now, we construct new programs $\bar{M}$ and
$\bar{M'}$ as follows.
\[
\begin{array}{cc}
  \bar{M}(h_1)=o_1 &\bar{M'}(h_1')=o_1'\\
  \bar{M}(h_2)=o_2 &\bar{M'}(h_2')=o_2'\\
  \dots&\dots\\
  \bar{M}(h_i)=o_i &\bar{M'}(h_j')=o_j'\\
  \bar{M}(h_{i+1})=o &\bar{M'}(h_{j+1}')=o_{j+1}'\\
  \bar{M}(h_{i+2})=o &\bar{M'}(h_{j+2}')=o_{j+2}'\\
  \dots&\dots\\
  \bar{M}(h_{i+j})=o &\bar{M'}(h_{j+i}')=o_{j+i}'\\
  \bar{M}(h_{i+j+1})=o_r & \bar{M'}(h_{j+i+1}')=o_r'\\
  \dots&\dots\\
  \bar{M}(h_{n})=o_r & \bar{M'}(h_{n}')=o_r'\\
\end{array}
\]
where 
\begin{itemize}
\item $o\not=o_r$,
\item $\aset{o_1,o_2,\dots,o_i}\cap\aset{o,o_r}=\emptyset$,
\item $o_{j+1}'$, $o_{j+2}'$, $\dots$, $o_{j+i}'$, and $o_r'$ are
  distinct,
\item
  $\aset{o_1',o_2',\dots,o_j'}\cap\aset{o_{j+1}',\dots,o_{j+i}',o_r'}=\emptyset$,
\item $\aset{h_1,\dots,h_n} = \aset{h_1',\dots,h_n'}$, and
\item $n=2k$.
\end{itemize}
Then, comparing the Shannon-entropy-based quantitative information flow
of $\bar{M}$ and $\bar{M'}$, we have, 
\[
\begin{array}{l}
  {\it SE}[U](\bar{M'})-{\it SE}[U](\bar{M})\\
  \hspace{2em}=\sum_{o_x'\in\aset{o_1',\dots,o_i'}}U(o_x')\log\frac{1}{U(o_x')}\\
  \hspace{3em}+U(o')\log\frac{1}{U(o')}+U(o_r')\log\frac{1}{U(o_r')}\\
  \hspace{4em}-(\sum_{o_x\in\aset{o_1,\dots,o_j}}U(o_x)\log\frac{1}{U(o_x)}\\
  \hspace{5em}+\sum_{o_y\in\aset{o_{j+1},\dots,o_{j+i}}}U(o_y)\log\frac{1}{U(o_y)}\\
  \hspace{6em}+U(o_r)\log\frac{1}{U(o_r)})\\
\end{array}
\]
(Note the abbreviations from Appendix~\ref{sec:lemdefs}.)
By lemma~\ref{lem:a7}, we have
\[
\begin{array}{l}
  \sum_{o_x\in\aset{o_1,\dots,o_i}}U(o_x)\log\frac{1}{U(o_x)}\\
  \qquad\qquad\le\sum_{o_y'\in\aset{o_{j+1}',\dots,o_{j+i}'}}U(o_y')\log\frac{1}{U(o_y')}
\end{array}
\]
and
\[
\begin{array}{l}
  U(o)\log\frac{1}{U(o)}
  \le\sum_{o_x'\in\aset{o_1',\dots,o_j'}}U(o_x')\log\frac{1}{U(o_x')}
\end{array}
\]
Trivially, we have
\[
U(o_r')\log\frac{1}{U(o_r')}=U(o_r)\log\frac{1}{U(o_r)}
\]
As a result, we have
\[
{\it SE}[U](\bar{M'})-{\it SE}[U](\bar{M})\ge 0
\]
Note that $\bar{M}$ and $\bar{M'}$ have the same counterexamples $T$
and $T'$, that is, $T\subseteq\sembrack{\bar{M}}$ and
$T'\subseteq\sembrack{\bar{M'}}$.  However, we have
$(\bar{M},\bar{M'})\in C_{\it SE}[U]$.  This leads to a contradiction.

\subsection{Complexities for Loop-free Boolean Programs}

\label{sec:complex}

\begin{figure}
\[
\begin{array}{rl}
M::=&x:=\psi\mid{\sf if}\; \psi \;{\sf then}\; M \;{\sf else}\; M \mid M_0 ; M_1\\
\phi,\psi::=&{\sf true}\mid x\mid \phi\wedge \psi\mid \neg \phi
\end{array}
\]
\caption{The syntax of loop-free boolean programs}\label{syntax}
\end{figure}

\begin{figure}
\[
\begin{array}{l}
\wpre{x:=\psi}{\phi}=\phi[\psi/x]\\
\wpre{{\sf if}\;\psi\;{\sf then}\;M_0\;{\sf else}\;M_1}{\phi}\\
\qquad=(\psi\Rightarrow \wpre{M_0}{\phi})\wedge(\neg \psi\Rightarrow \wpre{M_1}{\phi})\\
\wpre{M_0;M_1}{\phi}=\wpre{M_0}{\wpre{M_1}{\phi}}
\end{array}
\]
\caption{The weakest precondition for loop-free boolean programs}
\label{wpsemantics}
\end{figure}

The purpose of this section is to show a complexity theoretic gap
between non-interference and quantitative information flow.  The
results strengthen the hypothesis that quantitative information flow
is quite hard to compute precisely, and also suggest an interesting
connection to counting problems.

We focus on loop-free boolean programs whose syntax is given in
Figure~\ref{syntax}.  We assume the usual derived formulas $\phi
\Rightarrow \psi$, $\phi = \psi$, $\phi \vee \psi$, and ${\sf false}$.
We give the usual weakest precondition semantics in
Figure~\ref{wpsemantics}.

To adapt the information flow framework to boolean programs, we make
each information flow variable $H$, $L$, and $O$ range over functions
mapping boolean variables of its kind to boolean values.  So, for
example, if $x$ and $y$ are low security boolean variables and $z$ is
a high security boolean variable, then $L$ ranges over the functions
$\aset{x,y} \rightarrow \aset{{\sf false},{\sf true}}$, and $H$ and
$O$ range over $\aset{z} \rightarrow \aset{{\sf false},{\sf
true}}$.\footnote{ We do not distinguish input boolean variables from
output boolean variables.  But, a boolean variable can be made
output-only by assigning a constant to the variable at the start of
the program and made input-only by assigning a constant at the end.}
(Every boolean variable is either a low security boolean variable or a
high security boolean variable.)  We write $M(h,\ell) = o$ for an
input $(h,\ell)$ and an output $o$ if $(h,\ell) \models
\wpre{M}{\phi}$ for a boolean formula $\phi$ such that $o \models
\phi$ and $o' \not\models \phi$ for all output $o' \neq o$.  Here,
$\models$ is the usual logical satisfaction relation, using
$h,\ell,o$, etc.~to look up the values of the boolean variables.
(Note that this incurs two levels of lookup.)

As an example, consider the following program.
\[
\begin{array}{l}
M \equiv \\
\quad\ttassign{z}{x};\ttassign{w}{y};\\
\quad\ttif{x \wedge
  y}{\ttassign{z}{\neg z}}{\ttassign{w}{\neg w}}
\end{array}
\]
Let $x$, $y$ and $w$ be high security variables and $z$ be a low
security variable.  Then,
\[
\begin{array}{rcll}
{\it SE}[U](M) & = & 1.5\\

{\it ME}[U](M) & = & \log 3\\
&\approx & 1.5849625\\
{\it GE}[U](M)&=& 1.25\\
{\it CC}(M) & = & \log 3\\
&\approx & 1.5849625 \\
\end{array}
\]

We prove the following hardness results.  These results are proven by
a reduction from \#SAT, which is the problem of counting the number of
solutions to a quantifier-free boolean formula.  \#SAT is known to be
\#P-complete.  Because \#SAT is a function problem and the comparison
problems are decision problems, a step in the proofs makes binary
search queries to the comparison problem oracle a polynomial number of
times.  (Recall that the notation $\text{FP}^A$ means the complexity
class of function problems solvable in polynomial time with an oracle
for the problem $A$.)
\begin{theorem}\label{thm:secomp}
$\text{\#P}\subseteq \text{FP}^{C_{\it SE}[U]}$
\end{theorem}

\begin{theorem}\label{thm:mecomp}
$\text{\#P}\subseteq \text{FP}^{C_{\it ME}[U]}$
\end{theorem}

\begin{theorem}\label{thm:gecomp}
$\text{\#P}\subseteq \text{FP}^{C_{\it GE}[U]}$
\end{theorem}

\begin{theorem}\label{thm:cccomp}
$\text{\#P}\subseteq \text{FP}^{C_{\it CC}}$
\end{theorem}
We remind that the above results apply (even) when the comparison
problems $C_{\it SE}[U]$, $C_{\it ME}[U]$,  $C_{\it GE}[U]$, and $C_{\it CC}$ are
restricted to loop-free boolean programs.

In summary, each comparison problem $C_{\it SE}[U]$, $C_{\it ME}[U]$,
$C_{\it GE}[U]$, and $C_{\it CC}$ can be used a polynomial number of
times to solve a \#P-complete problem.  Because Toda's
theorem~\cite{toda91} implies that the entire polynomial hierarchy can
be solved by using a \#P-complete oracle a polynomial number of times,
our results show that the comparison problems for quantitative
information flow can also be used a polynomial number of times to
solve the entire polynomial hierarchy, for the case of loop-free
boolean programs.

As shown below, this presents a gap from non-interference, which is
only coNP-complete for loop-free boolean programs.
\begin{theorem}\label{thm:nicomp}
  Checking non-interference is coNP-complete for loop-free boolean programs.
\end{theorem}

The above is an instance of the general observation that, by solving
quantitative information flow problems, one is able to solve the class
of problems known as {\em counting problems},\footnote{Formally, a
counting problem is the problem of counting the number of solutions to
a decision problem.  For instance, \#P is the class of counting
problems associated with NP.} which coincides with \#SAT for the case
of loop-free boolean programs.

\subsection{Proof of Theorem~\ref{thm:secomp}}

We discuss the details of the proof of Theorem~\ref{thm:secomp}.  The
proofs of Theorems~\ref{thm:mecomp}, \ref{thm:gecomp}, \ref{thm:cccomp} are
deferred to Appendix~\ref{sec:proofs}.

First, we prove the following lemma which states that we can compare the
number of solutions to boolean formulas by computing ${\it SE}[U]$.
(For convenience, we use large letters $H$, $L$, $O$, etc.~to range
over boolean variables as well as generic random variables.)
\begin{lemma}
\label{lem:a3}
Let $\vect H$ and $H'$ be distinct boolean random variables.  Let $i$
and $j$ be any non-negative integers such that $i \le 2^{|\vect H|}$
and $j\le 2^{|\vect H|}$.  Let $\psi_i$ (resp. $\psi_j$) be a formula
over $\vect H$ having $i$ (resp. $j$) assignments.  Then, $j \leq i$
iff ${\it SE}[U](M_j)\le{\it SE}[U](M_i)$ where $M_j \equiv
O:=\psi_j\wedge H'$ and $M_i \equiv O:=\psi_i\wedge H'$.
\end{lemma}
\begin{IEEEproof}
Let $p=\frac{i}{2^{|H|+1}}$ and
$q=\frac{j}{2^{|H|+1}}$.  We have
\[
\begin{array}{rl}
  {\it SE}[U](M_j) & =\frac{j}{2^{|H|+1}}\log\frac{2^{|H|+1}}{j} + \frac{2^{|H|+1}-j}{2^{|H|+1}}\log\frac{2^{|H|+1}}{2^{|H|+1}-j}\\
\hspace{5.3em} & = p\log p + (1-p)\log \frac{1}{1-p}\\
  {\it SE}[U](M_i) & =\frac{i}{2^{|H|+1}}\log\frac{2^{|H|+1}}{i} + \frac{2^{|H|+1}-i}{2^{|H|+1}}\log\frac{2^{|H|+1}}{2^{|H|+1}-i}\\
\hspace{5.3em} & = q\log q + (1-q)\log \frac{1}{1-q}
\end{array}
\]

\begin{itemize}
\item {\bf Only If}

Suppose $j \leq i$.  Then,
\[
\begin{array}{l}
  {\it SE}[U](M_i)-{\it SE}[U](M_j) \\
\hspace{3em} = p\log\frac{1}{p} + (1-p)\log\frac{1}{1-p}\\
 \hspace{4em} - q\log\frac{1}{q} - (1-q)\log\frac{1}{1-q}\\
\hspace{3em} =\log(\frac{1-p}{p})^p \frac{1-q}{1-p} (\frac{q}{1-q})^q
\end{array}
\]
Then, from $\frac{1-q}{1-p}\ge 1$ and $p\ge q\ge 0$, we have
\[
\begin{array}{rl}
{\it SE}[U](M_i)-{\it SE}[U](M_j) & \geq \log(\frac{1-p}{p})^p (\frac{q}{1-q})^q\\
& \ge\log(\frac{1-p}{p})^q (\frac{q}{1-q})^q\\
& =\log(\frac{(1-p)q}{p(1-q)})^q\\
& =\log(\frac{q-pq}{p-pq})^q\\
& =\log(\frac{pq-q}{pq-p})^q\\
& =\log(\frac{1-\frac{1}{p}}{1-\frac{1}{q}})^q\\
& \geq 0
\end{array}
\]
The last line follows from $\frac{1-\frac{1}{p}}{1-\frac{1}{q}}\ge 1$.

\texcomment{
\[
\begin{array}{l}
  {\it SE}[U](M_i)-{\it SE}[U](M_j)\\
  \qquad=p\log\frac{1}{p} + (1-p)\log\frac{1}{1-p}\\
\qquad\qquad - q\log\frac{1}{q} - (1-q)\log\frac{1}{1-q}\\
  \qquad=\log(\frac{1}{p})^p + \log(\frac{1}{1-p})^{1-p} + \log q^q + \log (1-q)^{1-q}\\
  \qquad =\log(\frac{1}{p})^p (\frac{1}{1-p})^{1-p}q^q (1-q)^{1-q}\\
  \qquad =\log(\frac{1}{p})^p (1-p)^p \frac{1}{1-p} q^q (\frac{1}{1-q})^q(1-q)\\
  \qquad =\log(\frac{1-p}{p})^p \frac{1-q}{1-p} (\frac{q}{1-q})^q\\
  \qquad\ge\log(\frac{1-p}{p})^p (\frac{q}{1-q})^q\\
  \qquad\ge\log(\frac{1-p}{p})^q (\frac{q}{1-q})^q\\
  \qquad=\log(\frac{(1-p)q}{p(1-q)})^q\\
  \qquad=\log(\frac{q-pq}{p-pq})^q\\
  \qquad=\log(\frac{pq-q}{pq-p})^q\\
  \qquad=\log(\frac{1-\frac{1}{p}}{1-\frac{1}{q}})^q\\
  \qquad\ge 0
\end{array}
\]
}
\item {\bf If}

We prove the contraposition.  Suppose $j > i$.  Then, 
\[
\begin{array}{l}
{\it SE}[U](M_j) - {\it SE}[U](M_i) \\ \qquad = q\log\frac{1}{q} +
(1-q)\log\frac{1}{1-q} \\ \qquad\qquad- p\log\frac{1}{p} -
(1-p)\log\frac{1}{1-p} \\ \qquad > 0
\end{array}
\]
The last line follows from the fact that $0\le p < q\le \frac{1}{2}$.
Therefore, ${\it SE}[U](M_j) \not\leq {\it SE}[U](M_i)$.
\end{itemize}
\end{IEEEproof}

Then, using Lemma~\ref{lem:a3}, we prove the following lemma which is
crucial to proving Theorem~\ref{thm:secomp}.
\begin{lemma}
\label{lem:a4}
Let $\vect H$ be distinct variables and $\phi$ be a boolean formula
over $\vect H$.  Then, the number of assignments for $\phi$ can be
computed by executing an oracle that decides whether programs are in
$C_{\it SE}[U]$ at most $3*(|\vect H|+1)+2$ times.
\end{lemma}
\begin{IEEEproof}
First, we define a procedure that returns the number of solutions of
$\phi$.

Let $F(j) \equiv O:=\psi\wedge H'$ where $\psi$ is a formula over
$\vect H$ having $j$ assignments and $H'$ be a boolean variable such
that $H'\not\in \aset{\vect H}$.  Note that, by Lemma~\ref{lem:a2}, such
$\psi$ can be generated in linear time.

Then, we invoke the following procedure where $M'\equiv O':=\phi\wedge H'$.
\[
\begin{array}{l}
  l=0;\\
  r=2^{|\vect H|};\\
  n=(\ell+r)/2;\\
  {\sf while}\;\neg C_{\it SE}[U](F(n),M') \vee \neg C_{\it SE}[U](M',F(n))\\
  \qquad{\sf if}\;C_{\it SE}[U](F(n),M')\\
  \qquad\qquad{\sf then}\;\{\ell=n;n=(\ell+r)/2;\}\\
  \qquad\qquad{\sf else}\;\{r=n;n=(\ell+r)/2;\}\\
  {\sf return}\;n
\end{array}
\]

Note that when the procedure terminates, we have ${\it SE}[U](F(n))
= {\it SE}[U](M')$, and so by Lemma~\ref{lem:a3}, $n$ is the number of
satisfying assignments to $\phi$.  

We show that the procedure iterates at most $|\vect{H}|+1$ times.  To
see this, every iteration in the procedure narrows the range between
$r$ and $\ell$ by one half.  Because $r - \ell$ is bounded by
$2^{|\vect H|}$, it follows that the procedure iterates at most
$|\vect{H}|+1$ times.  Hence, the oracle $C_{\it SE}[U]$ is accessed
$3*(|\vect{H}|+1)+2$ times, and this proves the lemma.
\end{IEEEproof}

Finally, Theorem~\ref{thm:secomp} follows from Lemma~\ref{lem:a4} and
the fact that \#SAT, the problem of counting the number of solutions
to a boolean formula, is \#P-complete.

\section{Universally Quantifying Distributions}

\label{sec:qshan}

As proved in Section~\ref{sec:qifnksafe}, precisely computing
quantitative information flow is quite difficult.  Indeed, we have
shown that even just comparing two programs on which has the larger
flow is difficult (i.e., $C_{\it SE}$, $C_{\it ME}$, $C_{\it GE}$, and
$C_{\it CC}$).

In this section, we show that universally quantifying the
Shannon-entropy based comparison problem $C_{\it SE}[\mu]$, the
min-entropy based problem $C_{\it ME}[\mu]$, or the guessing-entropy
based problem $C_{\it GE}[\mu]$ over the distribution $\mu$ is
equivalent to a simple relation $R$ enjoying the following properties.
\begin{itemize}
\item[(1)] $R$ is a $2$-safety property.
\item[(2)] $R$ is coNP-complete for loop-free boolean programs.
\end{itemize}
Note that (1) implies that we can actually check if $(M_1, M_2) \in
C_{\it SE}[\mu]$ for all $\mu$ via self composition (and likewise for
$C_{\it ME}[\mu]$ and $C_{\it GE}[\mu]$).  We actually show in
Section~\ref{sec:qbysc} that we can even use the security-type-based
approach suggested by Terauchi and Aiken~\cite{terauchi:sas05} to
minimize code duplication during self composition (i.e., do {\em
  interleaved} self composition).

We remind that except for the coNP-completeness result
(Theorem~\ref{thm:rcomp}), the results in this section apply to any
(deterministic and terminating) programs and not just to loop-free
boolean programs.

\begin{definition}
\label{def:r}
We define $R$ to be the relation such that $R(M_1, M_2)$
iff for all $\ell \in \mathbb{L}$ and $h,h' \in \mathbb{H}$, if
$M_1(h,\ell) \not= M_1(h', \ell)$ then $M_2(h,\ell) \not= M_2(h',
\ell)$.  
\end{definition}

Note that $R(M_1,M_2)$ essentially says that if an attacker can
distinguish a pair of high security inputs by executing $M_1$, then
she could do the same by executing $M_2$.  Hence, $R$ naturally
expresses that $M_1$ is at least as secure as $M_2$.
\footnote{We note that notions similar to $R$ have appeared in
  literature (often in somewhat different
  representations)~\cite{DBLP:conf/isss2/SabelfeldM03,DBLP:conf/popl/LiZ05,DBLP:journals/logcom/ClarkHM05}.
  In particular, Clark et al.~\cite{DBLP:journals/logcom/ClarkHM05}
  have shown a result analogous to the $\subseteq$ direction of
  Theorem~\ref{thm:ReqSE} below.  But, $R$'s properties have not been
  fully investigated.}

It may be somewhat surprising that this simple relation is actually
equivalent to the rather complex entropy-based quantitative
information flow definitions when they are cast as comparison problems
and the distributions are universally quantified, as stated in the
following theorems.  First, we show that $R$ coincides exactly with
$C_{\it SE}$ with its distribution universally quantified.
\begin{theorem}
\label{thm:ReqSE}
$R = \aset{(M_1,M_2) \mid \forall \mu.C_{\it SE}[\mu](M_1,M_2)}$
\end{theorem}
The proof is detailed in Section~\ref{sec:reqseproof}.  The next two
theorems show that $R$ also coincides with $C_{\it ME}$ and $C_{\it
GE}$ with their distribution universally quantified.
\begin{theorem}
\label{thm:ReqME}
$R = \aset{(M_1,M_2) \mid \forall \mu.C_{\it ME}[\mu](M_1,M_2)}$
\end{theorem}

\begin{theorem}
\label{thm:ReqGE}
$R=\aset{(M_1,M_2)\mid\forall\mu.C_{\it GE}[\mu](M_1,M_2)}$
\end{theorem}
The first half of the $\subseteq$ direction of the proofs for the
theorems above is much like the that of Theorem~\ref{thm:ReqSE}, that
is, it makes the observation that $M_2$ disambiguates the high
security inputs at least as fine as does $M_1$.  Then, the proof
concludes by utilizing the particular mathematical properties relevant
to the respective definitions.  The proof for the $\supseteq$
direction is also similar to the argument used in
Theorem~\ref{thm:ReqSE}.  The details of the proofs appear in
Appendix~\ref{sec:proofs}.

Next, we show that $R$ refines $C_{\it CC}$ in the sense that if
$R(M_1, M_2)$ then $C_{\it CC}(M_1, M_2)$.
\begin{theorem}
\label{thm:RimpCC}
$R \subseteq C_{\it CC}$
\end{theorem}

Note that, the other direction, $R \supseteq C_{\it CC}$, does not
hold as $R$ is not always a total order, whereas $C_{\it CC}$ is.  We
also show that $R$ is compatible with the notion of non-interference.
\begin{theorem}
\label{thm:rni}
Let $M_2$ be a non-interferent program.  Then, $R(M_1, M_2)$ iff $M_1$
is also non-interferent and $M_1$ has the same input domain as $M_2$.
\end{theorem}

Next, we show that $R$ is easier to decide than the
non-universally-quantified versions of the comparison problems.
First, it is trivial to see from Definition~\ref{def:r} that $R$ is a
$2$-safety property.
\begin{theorem}
\label{thm:r2safe}
$R$ is a $2$-safety property.
\end{theorem}

It can be shown that, restricted to loop-free boolean programs, $R$ is
coNP-complete.  This follows directly from the observation that we can
decide $R$ by self composition thanks to its $2$-safety property and
the fact that, for loop-free boolean programs, self composition
reduces the problem to an UNSAT instance.\footnote{To construct a
  polynomial size boolean formula from a loop-free boolean program, we
  use the well-known efficient weakest precondition construction
  technique~\cite{DBLP:conf/popl/FlanaganS01,DBLP:journals/ipl/Leino05}
  instead of the naive rules given in Figure~\ref{wpsemantics}.}
\begin{theorem}\label{thm:rcomp}
Restricted to loop-free boolean programs, $R$ is coNP-complete.
\end{theorem}

\texcomment{
As an aside, we note that existentially quantifying the distribution
$\mu$ in the comparison problem $C_{\it SE}[\mu]$ or $C_{\it ME}[\mu]$
does not lead to a meaningful notion, because for any $M_1$ and $M_2$
with the same input domain, $\exists\mu.{\it SE}[\mu](M_1) \leq {\it
  SE}[\mu](M_2)$, $\exists\mu.{\it ME}[\mu](M_1) \leq {\it
  ME}[\mu](M_2)$, and  $\exists\mu.{\it GE}[\mu](M_1) \leq {\it
  GE}[\mu](M_2)$ .
}

\subsection{Proof of Theorem~\ref{thm:ReqSE}}

\label{sec:reqseproof}

We discuss the details of the proof of Theorem~\ref{thm:ReqSE}.  The
proofs of Theorems~\ref{thm:ReqME}, \ref{thm:ReqGE}, \ref{thm:RimpCC} are
deferred to Appendix~\ref{sec:proofs}.

First, we prove the following lemma which says that, if $R(M,M')$ then
${\it SE}[U](M')$ is at least as large as ${\it SE}[U](M)$ per each
low security input $\ell \in \mathbb{L}$.
\begin{lemma}
\label{lem:a8}
Suppose $R(M,M')$, that is, for all $h_1$, $h_2$ in $\mathbb{H}$ and
$\ell$ in $\mathbb{L}$, $M'(h_1,\ell)= M'(h_2,\ell)\Rightarrow
M(h_1,\ell)= M(h_2,\ell)$.  Let $\mathbb{O}$ be the set of the outputs
of $M$, and $\mathbb{O'}$ be the set of the outputs of $M'$.  Then,
for any $\ell$, we have $\sum_{o\in\mathbb{O}}
\mu(o,\ell)\log\frac{\mu(\ell)}{\mu(o,\ell)}\le\sum_{o'\in\mathbb{O'}}
\mu(o',\ell)\log\frac{\mu(\ell)}{\mu(o',\ell)}$.  (Recall the
notational convention from Definition~\ref{def:distabrv}.)
\end{lemma}
\begin{IEEEproof}
  First, we prove for any output $o$ of $M$, there exist corresponding
  outputs $\mathbb{O}_o=\aset{o_0',o_1',\dots,o_n'}$ of $M'$ such that
\[
\begin{array}{l}\mu(o,\ell)\log\frac{\mu(\ell)}{\mu(o,\ell)}\\
\ \ \le \sum_{o_r'\in\mathbb{O}_o}
\mu(o_r',\ell)\log\frac{\mu(\ell)}{\mu(o_r',\ell)}
\end{array}
\]
Let $\mathbb{H}_o$ be the set such that $\mathbb{H}_o=\aset{h\mid
  M(h,\ell)=o}$.  Let $\aset{h_0,h_1,\dots,h_n}=\mathbb{H}_o$.  Let
$o_0'=M'(h_0,\ell)$,\dots and, $o_n'=M'(h_n,\ell)$.  For any $h'$ such
that $o_r'=M'(h',\ell)$ and $o_r'\in\aset{o_0',o_1',\dots,o_n'}$, we
have $h'\in\aset{h_1,\dots, h_n}$ since $R(M,M')$.  Then, we have
$\mu(o,\ell) = \sum_{o_r'\in\aset{o_1',\dots,o_n'}} \mu(o_r',\ell)$.
By Lemma~\ref{lem:a7}, we have
\[
\begin{array}{l}
  \mu(o,\ell)\log\frac{\mu(\ell)}{\mu(o,\ell)}\\
  \ \ \le \sum_{o_r'\in\aset{o_0',o_1',\dots,o_n'}}
  \mu(o_r',\ell)\log\frac{\mu(\ell)}{\mu(o_r',\ell)}
\end{array}
\]
Now to prove the lemma, it suffices to show that each $\mathbb{O}_o$
constructed above are disjoint.  That is, for $o_1$ and $o_2$ outputs
of $M$ such that $o_1\not=o_2$, $\mathbb{O}_{o_1}\cap
\mathbb{O}_{o_2}=\emptyset$.  For contradiction, suppose
$o'\in\mathbb{O}_{o_1}\cap \mathbb{O}_{o_2}$.  Then, there exist $h_1$
and $h_2$ such that $o_1=M(h_1,\ell)$, $o'=M'(h_1,\ell)$,
$o_2=M(h_2,\ell)$, and $o'=M'(h_2,\ell)$.  Since $R(M,M')$, we have
$o_1=o_2$, and it leads to a contradiction.  Hence, we have
\[
\sum_o \mu(o,\ell)\log\frac{\mu(\ell)}{\mu(o,\ell)}\le\sum_{o'}\mu(o',\ell)\log\frac{\mu(\ell)}{\mu(o',\ell)}
\]
\end{IEEEproof}

We now prove Theorem~\ref{thm:ReqSE}.
\begin{IEEEproof}

\begin{itemize}
\item $\subseteq$

  Suppose $(M,M')\in R$.  By Lemma~\ref{lem:detse},
\[
\begin{array}{rl}
{\it SE}[\mu](M)&=\mathcal H[\mu](O|L)\\
  &=\sum_\ell\sum_o \mu(o,\ell)\log\frac{\mu(\ell)}{\mu(o,\ell)}\\
\end{array}
\]
and 
\[
\begin{array}{rl}
{\it SE}[\mu](M')&=\mathcal H[\mu](O'|L)\\
  &=\sum_\ell\sum_{o'} \mu(o',\ell)\log\frac{\mu(\ell)}{\mu(o',\ell)}\\
\end{array}
\]
By Lemma~\ref{lem:a8} and the fact that $(M,M')\in R$, we obtain for
any $\ell$
\[
\sum_o \mu(o,\ell)\log\frac{\mu(\ell)}{\mu(o,\ell)}\le \sum_{o'} \mu(o',\ell)\log\frac{\mu(\ell)}{\mu(o',\ell)}
\]
Hence,
\[
\begin{array}{l}
\sum_\ell\sum_o \mu(o,\ell)\log\frac{\mu(\ell)}{\mu(o,\ell)}\\
\qquad\le \sum_\ell\sum_{o'} \mu(o',\ell)\log\frac{\mu(\ell)}{\mu(o',\ell)}
\end{array}
\]
\item $\supseteq$

  We prove the contraposition.  Suppose $(M,M')\not\in R$.  Then,
  there exist $o',h_0,h_1,\ell'$ such that
  $o'=M'(h_0,\ell')=M'(h_1,\ell')$ and $M(h_0,\ell')\not=
  M(h_1,\ell')$.  Pick a probability function $\mu$ such that
  $\mu(h_0,\ell')=\mu(h_1,\ell')=\frac{1}{2}$.

Then, we have
\[
\begin{array}{rl}
H[\mu](O'|L)&=\sum_\ell\sum_o \mu(o,\ell)\log\frac{\mu(\ell)}{\mu(o,\ell)}\\
&= \mu(o',\ell')\log\frac{\mu(\ell')}{\mu(o',\ell')}\\
&= 1\log\frac{1}{1}\\
&= 0
\end{array}
\]
Let $o_0$ and $o_1$ be output variables such that $o_0=M(h_0,\ell')$,
$o_1=M(h_1,\ell')$, and $o_0\not=o_1$.
\[
\begin{array}{rl}
\mathcal{H}[\mu](O|L)&=\sum_{o\in\aset{o_0,o_1}} \mu(o,\ell')\log\frac{\mu(\ell')}{\mu(o,\ell')}\\
&=\frac{1}{2}\log\frac{1}{\frac{1}{2}}+\frac{1}{2}\log\frac{1}{\frac{1}{2}}\\
&=1\\
\end{array}
\]
Therefore, ${\it SE}[\mu](M)\not\le{\it SE}[\mu](M')$, that is,
$(M,M')\not\in\aset{(M_1,M_2)\mid\forall\mu.(M_1,M_2) \in C_{\it
    SE}[\mu]}$.
\end{itemize}
\end{IEEEproof}

\subsection{Quantitative Information Flow via Self Composition}

\label{sec:qbysc}

Theorems~\ref{thm:ReqSE}, \ref{thm:ReqME}, \ref{thm:ReqGE}, and
\ref{thm:r2safe} imply that we can check if the entropy-based
quantitative information flow of a program (i.e., {\it SE}, {\it ME},
and {\it GE}) is bounded by that of another for all distributions via
self composition~\cite{barthe:csfw04,darvas:spc05}.  This suggests a
novel approach to precisely checking quantitative information flow.

That is, given a {\em target} program $M_1$, the user would construct
a {\em specification} program $M_2$ with the same input domain as
$M_1$ having the desired level of security.  Then, she would check
$R(M_1, M_2)$ via self composition.  If so, then $M_1$ is guaranteed
to be at least as secure as $M_2$ according to the Shannon-entropy
based, the min-entropy based, and the guessing-entropy based
definition of quantitative information flow for all distributions (and
also channel-capacity based definition), and otherwise, there must be
a distribution in which $M_1$ is less secure than $M_2$ according to
the entropy-based definitions.

Note that deciding $R(M_1,M_2)$ is useful even when $M_1$ and $M_2$
are $R$-incomparable, that is, when neither $R(M_1, M_2)$ nor $R(M_2,
M_1)$.  This is because $\neg R(M_1, M_2)$ implies that $M_1$ is less
secure than $M_2$ on some distribution.

For example, suppose $M_1$ is some complex login program with the high
security input $H$ and the low security input $L$.  And we would like
to verify that $M_1$ is at least as secure as the prototypical login
program $M_2$ below.
\[
M_2 \equiv \ttif{H = L}{\ttassign{O}{0}}{\ttassign{O}{1}}
\]
Then, using this framework, it suffices to just query if $R(M_1, M_2)$
is true.  (Note that the output domains of $M_1$ and $M_2$ need not to
match.)

We now describe how to actually check $R(M_1, M_2)$ via
self composition.  From $M_1$ and $M_2$, we construct the
self-composed program $M'$ shown below.
\[
\begin{array}{l}
M'(H,H',L) \equiv  \\
\hspace{1em}O_1 := M_1(H,L);O_1' := M_1(H',L);\; //\: {\tt L1}\\
\hspace{1em}O_2 := M_2(H,L);O_2' := M_2(H',L);\; //\: {\tt L2}\\
\hspace{1em} {\tt assert}( O_1 \neq O_1' \Rightarrow O_2 \neq O_2' )
\end{array}
\]
Note that $R(M_1, M_2)$ is true iff $M'$ does not cause an assertion
failure.  The latter can be checked via a software safety verifier
such as SLAM and
BLAST~\cite{DBLP:conf/popl/BallR02,DBLP:conf/popl/HenzingerJMS02,mcmillan:cav06,DBLP:journals/sttt/BeyerHJM07}.
As an aside, we note that this kind of construction could be easily
generalized to reduce any $k$-safety problem
(cf. Section~\ref{sec:qifnksafe}) to a safety problem, as shown by
Clarkson and Schneider~\cite{DBLP:conf/csfw/ClarksonS08}.

Note that the line ${\tt L1}$ (resp.~${\tt L2}$) of the pseudo code
above is $M_1$ (resp.~$M_2$) sequentially composed with a copy of
itself, which is from where the name ``self composition'' comes.
Therefore, technically, $M'$ is a composition of two self
compositions.

${\tt L1}$ (and ${\tt L2}$) are actually exactly the original self
composition proposed for
non-interference~\cite{barthe:csfw04,darvas:spc05}.  Terauchi and
Aiken~\cite{terauchi:sas05} noted that only the parts of $M_1$ (and
$M_2$) that depend on the high security inputs $H$ and $H'$ need to be
duplicated and self composed, with the rest of the program left intact
and ``interleaved'' with the self-composed parts.  The resulting
program tends to be verified easier than the naive self composition by
modern software safety verifiers.

They proposed a set of transformation rules that translates a WHILE
program annotated with security types~\cite{volpano96sound} (or
dependency analysis results) to an interleaved self-composed program.
This was subsequently improved by a number of researchers to support a
richer set of language features and transformation
patterns~\cite{unno:plas2006,naumann:esorics06}.  These transformation
methods can be used in place of the naive self compositions at ${\tt
  L1}$ and ${\tt L2}$ in building $M'$.  That is, we apply a security
type inference (or a dependency analysis) to $M_1$ and $M_2$ to infer
program parts that depend on the high security inputs $H$ and $H'$ so
as to only duplicate and self compose those parts of $M_1$ and $M_2$.

\subsection{Example}

\label{sec:example}

\indent
We recall the ideal login program below.
\[
M_{\it spec} \equiv \ttif{H = L}{\ttassign{O}{0}}{\ttassign{O}{1}}
\]
We check the following four programs using the above as the
specification.
\[
\begin{array}{l}
M_1 \equiv \ttassign{O}{H} \\\\
M_2 \equiv \ttif{H = L}{\ttassign{O}{0}}{\ttassign{O}{H \texttt{\&} 1}} \\\\
M_3 \equiv \ttassign{O}{1};\ttassign{i}{0};\\
\hspace{2.8em}{\tt while}\; i < 32\;\{ \\
\hspace{3.5em}\ttassign{m}{1\;\texttt{<<}\;i};\\
\hspace{3.5em}\texttt{if}\;H \texttt{\&}m \neq  L \texttt{\&}m\;\texttt{then}\\
\hspace{4.2em}\ttassign{O}{0};{\sf break};\\
\hspace{3.5em}\texttt{else}\\
\hspace{4.2em}i\texttt{++};\\
\hspace{2.8em}\}\\\\
M_4 \equiv \ttassign{O}{1};\ttassign{i}{0};\\
\hspace{2.8em}{\tt while}\; i < 64\;\{ \\
\hspace{3.5em}\ttassign{m}{1\;\texttt{<<}\;i};\\
\hspace{3.5em}\texttt{if}\;H \texttt{\&}m \neq  L \texttt{\&}m\;\texttt{then}\\
\hspace{4.2em}\ttassign{O}{0};{\sf break};\\
\hspace{3.5em}\texttt{else}\\
\hspace{4.2em}i\texttt{++};\\
\hspace{2.8em}\}
\end{array}
\]
Here, $H$ and $L$ are 64-bit values, $\texttt{\&}$ is the bit-wise and
operator, and $\texttt{<<}$ is the left shift operator.  $M_1$ leaks
the entire password.  $M_2$ checks the password against the user
guess but then leaks the first bit when the check fails.  $M_3$ only
checks the first 32 bits of the password.  And, $M_4$ implements password
checking correctly via a while loop.

We verify that only $M_4$ satisfies the specification, that is,
$R(M_4, M_{\it spec})$.  To see that $\neg R(M_1, M_{\it spec})$, note
that for any $\ell, h, h'$ such that $h \neq \ell$, $h' \neq \ell$ and
$h \neq h'$, we have that $M_1(h,\ell) \neq M_1(h',\ell)$ but $M_{\it
  spec}(h,\ell) = M_{\it spec}(h',\ell) = 1$.  To see that $\neg
R(M_2, M_{\it spec})$, note that for $\ell, h, h'$ such that $h \neq
\ell$, $h' \neq \ell$, $h \texttt{\&} 1 = 1$ and $h' \texttt{\&} 1 =
0$, we have that $1 = M_2(h,\ell) \neq M_2(h',\ell) = 0$ but $M_{\it
  spec}(h,\ell) = M_{\it spec}(h',\ell) = 1$.  To see that $\neg
R(M_3, M_{\it spec})$, let $\ell, h, h'$ be such that $h|_{32} =
\ell|_{32}$, $h'|_{32} \neq \ell|_{32}$, and $h \neq \ell$, then, $1 =
M_3(h,\ell) \neq M_3(h',\ell) = 0$ but $M_{\it spec}(h,\ell) = M_{\it
  spec}(h',\ell) = 1$.\footnote{It can be also shown that $\neg
  R(M_{\it spec}, M_2)$ and $\neg R(M_{\it spec}, M_3)$, that is,
  $M_2$ and $M_3$ are $R$-incomparable with $M_{\it spec}$.}  (Here,
$x|_{32}$ denotes $x\:\text{mod}\:2^{32}$, i.e., the first 32 bits of
$x$.)

The results imply that for $M_1$, $M_2$, and $M_3$, there must be a
distribution where the program is less secure than $M_{\it spec}$
according to each of the entropy-based definition of quantitative
information flow.  For instance, for the Shannon-entropy based
definition, we have for the uniform distribution $U$,
\[
\begin{array}{rl}
{\it SE}[U](M_{\it spec})&=\frac{1}{2^{58}} + \frac{2^{64}-1}{2^{64}}\log\frac{2^{64}}{2^{64}-1}\\
&\approx 3.46944695 \times 10^{-18}\\
{\it SE}[U](M_1)&=64\\
{\it SE}[U](M_2)&= \frac{1}{2} + \frac{1+2^{63}}{2^{65}}\log\frac{2^{64}}{1+2^{63}}+\frac{2^{63}-1}{2^{65}}\log\frac{2^{64}}{2^{63}-1}\\
&\approx 1.0\\
{\it SE}[U](M_3)&=\frac{1}{2^{27}}+\frac{2^{64}-2^{32}}{2^{64}}\log\frac{2^{64}}{2^{64}-2^{32}}\\
&\approx 7.78648\times10^{-9}\\
\end{array}
\]
That is, ${\it SE}[U](M_1) \not\leq {\it SE}[U](M_{\it spec})$, ${\it
  SE}[U](M_2) \not\leq {\it SE}[U](M_{\it spec})$, and ${\it
  SE}[U](M_3) \not\leq {\it SE}[U](M_{\it spec})$.  \texcomment{We note that the
same distribution does not always work as a counterexample for
different definitions.  For instance, ${\it ME}[U](M_2) = {\it
  ME}[U](M_3) = {\it ME}[U](M_{\it spec}) = 1$.}

Finally, we have that $R(M_4, M_{\it spec})$, and so $M_4$ is at least
as secure as $M_{\it spec}$ according to all of the definitions of
quantitative information flow considered in this paper.  In fact, it
can be also shown that $R(M_{\it spec}, M_4)$. (However, note that
$M_4$ and $M_{\it spec}$ are not semantically equivalent, i.e., their
outputs are reversed.)

\texcomment{
\[
\begin{array}{rl}
{\it SE}[U](M_{\it spec})&=\frac{1}{2^{64}}\log 2^{64} + \frac{2^{64}-1}{2^{64}}\log\frac{2^{64}}{2^{64}-1}\\
&\approx 3.46944695 \times 10^{-18}\\
{\it ME}[U](M_{\it spec})&=1\\
{\it GE}[U](M_{\it spec})&=\frac{1}{2}(2^{64}+1)\\
&\quad -(\frac{1}{2^{64\times 2}}\\
{\it CC}(M_{\it spec})&=1\\\\
{\it SE}[U](M_1)&=64\\
{\it ME}[U](M_1)&=64\\
{\it GE}[U](M_1)&=\frac{1}{2}(2^{64}+1)-\\
&=\\
{\it CC}(M_1)&=64\\\\
{\it SE}[U](M_2)&= \frac{1}{2} + 2^{63}(\frac{1+2^{63}}{2^{64\times 2}}\log\frac{2^{64}}{1+2^{63}}+\frac{2^{63}-1}{2^{64\times 2}}\log\frac{2^{64}}{2^{63}-1})\\
& \approx 1.0\\

\texcomment {2^{63}(\frac{2^{63}}{2^{64\times 2}}\log\frac{2^{64}}{2^{63}}+\frac{2^{63}}{2^{64\times 2}}\log\frac{2^{64}}{2^{63}}\\
&\quad + \frac{1+2^{63}}{2^{64\times 2}}\log\frac{2^{64}}{1+2^{63}}+\frac{2^{63}-1}{2^{64\times 2}}\log\frac{2^{64}}{2^{63}-1})\\
}

{\it ME}[U](M_2)&=\log\frac{3}{2}\\
&\approx 0.584963\\
{\it GE}[U](M_2)&=\frac{1}{2}(2^{64}+1)-\\
{\it CC}(M_2)&=1\\\\
{\it SE}[U](M_3)&=\frac{2^{32}}{2^{64}}\log\frac{2^{64}}{2^{32}}+\frac{2^{64}-2^{32}}{2^{64}}\log\frac{2^{64}}{2^{64}-2^{32}}\\
{\it ME}[U](M_3)&=1\\
{\it GE}[U](M_3)&=\\
{\it CC}(M_3)&=1\\\\
{\it SE}[U](M_4)&=\frac{1}{2^{64}}\log 2^{64} + \frac{2^{64}-1}{2^{64}}\log\frac{2^{64}}{2^{64}-1}\\
{\it ME}[U](M_4)&=1\\
{\it GE}[U](M_4)&=\\
{\it CC}(M_4)&=1\\

\end{array}
\]
}

\section{Related Work}

\label{sec:related}

This work builds on previous work that proposed information theoretic
notions of quantitative information
flow~\cite{denning82,clarkjcs2007,malacaria:popl2007,smith09,kopf07,DBLP:conf/sp/BackesKR09,mccamant:pldi2008,malacaria08,NMS2009}.
The previous research has mostly focused on information theoretic
properties of the definitions and proposed approximate (i.e.,
incomplete and/or unsound) methods for checking and inferring them.
In contrast, this paper investigates the verification theoretic and
complexity theoretic hardness of precisely inferring quantitative
information flow according to the definitions and also proposes a
precise method for checking quantitative information flow.  Our method
checks the quantitative information flow of a program against that of
a specification program having the desired level of security via self
composition for all distributions according to the entropy-based
definitions.

It is quite interesting that the relation $R$ unifies the different
proposals for the definition of quantitative information flow when
they are cast as comparison problems and their distributions are
universally quantified.  As remarked in Section~\ref{sec:qshan}, $R$
naturally expresses the fact that one program is more secure than the
other, and it could be argued that it is the essence of quantitative
information flow.  \texcomment{As shown in Section~\ref{sec:example},
  the unification property does not hold if we compared the programs
  with a fixed distribution such as the uniform distribution.}

Researchers have also proposed definitions of quantitative information
flow that do not fit the models studied in this paper.  These include
the definition based on the notion of {\em
  belief}~\cite{clarkson:csf2005}, and the ones that take the maximum
over the low security
inputs~\cite{malacaria:popl2007,kopf07}.\footnote{It is actually
  possible to show that the relation $R$ refines these notions in the
  same sense as Theorem~\ref{thm:RimpCC}, but the other direction is
  not guaranteed to hold.}

Despite the staggering complexity made apparent in this paper, recent
attempts have been made to (more) precisely infer quantitative
information flow (without universally quantifying over the
distribution as in our approach).  These methods are based on the idea
of {\em counting}.  As remarked in Section~\ref{sec:complex},
quantitative information flow is closely related to counting problems,
and several attempts have been made to reduce quantitative information
flow problems to them.\footnote{Note that our results only show that,
  {\em restricted to loop-free boolean programs}, the comparison
  problems can be {\em reduced from} \#SAT, and they do not show how
  to reduce them (or more general cases) {\em to} \#SAT or other
  counting problems.}  For instance, Newsome et al.~\cite{NMS2009}
reduce the inference problem to the \#SAT problem and apply
off-the-shelf \#SAT solvers.  To achieve scalability, they sacrifice
both soundness and completeness by only computing information flow
from one execution path.  Backes et al.~\cite{DBLP:conf/sp/BackesKR09}
also propose a counting-based approach that involves self composition.
However, unlike our method, they use self composition repeatedly to
find a new solution (i.e., more than a bounded number of times), and
so their results do not contradict the negative results of this paper.

\section{Conclusion}

\label{sec:concl}

We have investigated the hardness and possibilities of precisely
checking and inferring quantitative information flow according to the
various definitions proposed in literature.  Specifically, we have
considered the definitions based on the Shannon entropy, the min
entropy, the guessing entropy, and channel capacity.

We have shown that comparing two programs on which has the larger flow
according to these definitions is not a $k$-safety problem for any
$k$, and therefore that it is not possible to reduce the problem to a
safety problem via self composition.  The result is in contrast to
non-interference which is a $2$-safety problem.  We have also shown a
complexity theoretic gap with non-interference by proving the
\#P-hardness of the comparison problems and coNP-completeness of
non-interference, when restricted to loop-free boolean programs.

We have also shown a positive result that checking if the
entropy-based quantitative information flow of one program is larger
than that of another for all distributions is a $2$-safety problem,
and that it is also coNP-complete when restricted to loop-free boolean
programs.

We have done this by proving a surprising result that universally
quantifying the distribution in the comparison problem for the
entropy-based definitions is equivalent to a simple $2$-safety
relation.\texcomment{$R$ naturally expresses the property that one
  program is more secure than the other and also refines channel
  capacity.}  Motivated by the result, we have proposed a novel
approach to precisely checking quantitative information flow that
reduces the problem to a safety problem via self composition.  Our
method checks the quantitative information flow of a program for all
distributions against that of a specification program having the
desired level of security.

\texcomment{
An interesting future research direction is to see if the counting
based approaches can be combined with the self composition method
proposed in this paper to solve quantitative information flow
problems.
}

\section*{Acknowledgment}
We would like to thank Takeshi Tsukada for important insights and
useful discussions that motivated this work.  We also thank the
anonymous reviewers for their useful comments.  This work was
supported by MEXT KAKENHI 20700019 and 20240001, and Global COE
Program ``CERIES.''

\appendices

\section{Supporting Definitions and Lemmas}

\label{sec:lemdefs}

We define some abbreviations.
\begin{definition}
\label{def:distabrv}
  $\mu(x)\triangleq \mu(X=x)$
\end{definition}
We use this notation whenever the correspondences between random variables
and their values are clear.

For convenience, we sometimes use large letters $H$, $L$, $O$, etc.~to range
over boolean variables as well as generic random variables.

For simplicity, we often compute the Shannon-entropy based quantitative
information flow for programs that do not have low security inputs.
For such programs, the equation {\it SE} from Definition~\ref{def:se}
can be simplified as follows.
\begin{lemma}
\[
\begin{array}{rl}
  {\it SE}[\mu](M)&=\mathcal{I}[\mu](O;H)\\
  &=\mathcal{H}[\mu](O)
\end{array}
\]
\end{lemma}

We note the following property of deterministic programs~\cite{clark05}.
\begin{lemma}
\label{lem:detse}
For $M$ deterministic, 
\[
{\it SE}[\mu](M)=\mathcal{I}[\mu](O;H|L) = \mathcal{H}[\mu](O|L)
\]
\end{lemma}

The following lemma is used to show that we can generate a boolean
formula that has exactly the desired number of solutions in polynomial
(actually, linear) time.
\begin{lemma}
\label{lem:a2}
Let $k$ be an integer such that $0\le k \le 2^{|\vect x|}-1$.  Then, a
boolean formula that has exactly $k$ assignments over the variables
$\vect x$ can be computed in time linear in $|\vect x|$.
\end{lemma}
\begin{IEEEproof}
We define a procedure \textrm{iter} that returns the boolean formula.
Below, $\vect{x} = x_1,x_2,\dots$, i.e., $x_i$ is the $i$th variable.
\[
\begin{array}{rl}
\textrm{iter}(\epsilon,0)&={\sf false}\\
\textrm{iter}(0\ell,i)&=x_i\wedge(\textrm{iter}(\ell,i-1))\\
\textrm{iter}(1\ell,i)&=x_i\vee(\textrm{iter}(\ell,i-1))\\
\end{array}
\]
Here, $\epsilon$ is an empty string.  Let $\ell_k$ be a $|\vect
x|$-bit binary representation of $k$.  We prove that
$\textrm{iter}(\ell_k,|\vect x|)$ returns a boolean formula that has
exactly k assignments by induction on the number of variables, that
is, $|\vect x|$.

\begin{itemize}
\item $|\vect x|=1$
\begin{itemize}
\item $k=0$

$\textrm{iter}(0,1)$ returns $x_1\wedge {\sf false}$, that is, ${\sf
false}$.  ${\sf false}$ has no satisfying assignment.
\item $k=1$

$\textrm{iter}(1,1)$ returns $x_1\vee {\sf false}$, that is, $x_1$.
$x_1$ has only one satisfying assignment.
\end{itemize}

\item $|\vect x,x'|$
\begin{itemize}
\item $k < 2^{|\vect x,x'|-1}$

Let $0\ell$ be a binary representation of $k$.
$\textrm{iter}(0\ell,|\vect x,x'|)$ returns
$x'\wedge\textrm{iter}(\ell,|\vect x|)$.  By induction hypothesis,
$\textrm{iter}(\ell,|\vect x|)$ has $k$ satisfying assignments for
$\vect x$.  It follows that $x'\wedge\textrm{iter}(\ell,|\vect x|)$
has just $k$ satisfying assignments, because ${\sf
false}\wedge\textrm{iter}(\ell,|\vect x|)$ has no assignment and ${\sf
true}\wedge\textrm{iter}(\ell,|\vect x|)$ has just $k$ assignments.
\item $k\ge 2^{|\vect x|}$ 

Let $1\ell$ be a binary representation of $k$.
$\textrm{iter}(1\ell,|\vect x,x'|)$ returns
$x'\vee\textrm{iter}(\ell,|\vect x|)$.  $\ell$ is a binary
representation of $k-2^{|\vect x|}$.  By induction hypothesis,
$\textrm{iter}(\ell,|\vect x|)$ has $k-2^{|\vect x|}$ satisfying
assignments for $\vect x$.  It follows that
$x'\vee\textrm{iter}(\ell,|\vect x|)$ has just $k$ satisfying
assignments, because ${\sf false}\vee\textrm{iter}(\ell,|\vect x|)$
has just $k-2^{|\vect x|}$ assignments and when $x'={\sf true}$,
$x'\vee\textrm{iter}(\ell,|\vect x|)$ has just $2^{|\vect x|}$
assignments.
\end{itemize}
\end{itemize}
\end{IEEEproof}

We frequent the following property of logarithmic arithmetic when
proving statements concerning the Shannon entropy.
\begin{lemma}
\label{lem:a7}
Let $p$ and $q$ be numbers such that $p,q\in [0,1]$.  Then, we have
$p\log\frac{1}{p} + q\log\frac{1}{q}\ge (p+q)\log\frac{1}{p+q}$.
\end{lemma}
\begin{IEEEproof}
Because $\frac{p+q}{p}\ge 1$ and $\frac{p+q}{q}\ge 1$, it follows that,
\[
\begin{array}{l}
p\log\frac{1}{p} + q\log\frac{1}{q}-(p+q)\log\frac{1}{p+q}\\
\quad=p\log\frac{1}{p}-p\log\frac{1}{p+q}+q\log\frac{1}{q}-q\log\frac{1}{p+q}\\
\quad=p\log\frac{p+q}{p}+q\log\frac{p+q}{q}\\
\quad\ge 0
\end{array}
\]
\end{IEEEproof}

\section{Omitted Proofs}

\label{sec:proofs}

\begin{reftheorem}{\ref{thm:ReqNI}}
  Let $M$ be a program that takes high-security input $H$, low-security
  input $L$, and returns low-security output $O$.  Then, $M$ is non-interferent if
  and only if $\forall \mu.{\it SE}[\mu](M)=0$.
\end{reftheorem}
\begin{IEEEproof}
Recall that $M$ is non-interferent iff for any $h,h'\in \mathbb{H}$
and $\ell\in\mathbb{L}$, $M(h,\ell)=M(h',\ell)$.

\begin{itemize}
\item ($\Rightarrow$)
Suppose that $M$ is non-interferent.  Then, by Lemma~\ref{lem:detse},
\[
\begin{array}{rcl}
{\it SE}[\mu](M)&=&\mathcal{I}[\mu](O;H|L)\\
&=&\mathcal{H}[\mu](O|L)\\
&=&\sum_{o}\sum_{\ell}\mu(o,\ell)\log\frac{\mu(\ell)}{\mu(o,\ell)}\\
&=&\sum_{o}\sum_{\ell}\mu(o,\ell)\log\frac{\mu(\ell)}{\mu(\ell)}\\
&=&0
\end{array}
\]
The last step follows from the fact that non-interference implies
$\mu(\ell)=\mu(o,\ell)$.
\item ($\Leftarrow$) Suppose that $M$ is
  interferent.   Then, there must be $h_0$ and $h_1$ such that
 $M(h_0,\ell')=o_0$, $M(h_1,\ell')=o_1$, and $o_0\not=o_1$.
Pick a probability function $\mu$ such that 
$\mu(h_0,\ell')=\mu(h_1,\ell')=\frac{1}{2}$.
Then, by Lemma~\ref{lem:detse},
\[
\begin{array}{rcl}
{\it SE}[\mu](M)&=&\mathcal{I}[\mu](O;H|L)\\
&=&\mathcal{H}[\mu](O|L)\\
&=&\sum_{o}\sum_{\ell}\mu(o,\ell)\log\frac{\mu(\ell)}{\mu(o,\ell)}\\
&=&\mu(o_0,\ell')\log\frac{\mu(\ell')}{\mu(o_0,\ell')}\\
&&\qquad+\mu(o_1,\ell')\log\frac{\mu(\ell')}{\mu(o_1,\ell')}\\
&=&\frac{1}{2}\log 2 + \frac{1}{2}\log 2\\
&=&1
\end{array}
\]
Therefore, there exists $\mu$ such that ${\it SE}[\mu](M) \neq 0$,
and we have the conclusion.
\end{itemize}
\end{IEEEproof}

We note the following equivalence of {\it CC} and {\it ME}[U] for programs
without low security inputs~\cite{smith09}.
\begin{lemma}
\label{lem:ccme}
Let $M$ be a program without low security input.  Then, ${\it
  ME}[U](M) = {\it CC}(M)$.
\end{lemma}

The min-entropy-based quantitative information flow with uniformly
distributed high security input has the following property~\cite{smith09}.
\begin{lemma}
\label{lem:melog}
Let $M$ be a program without low security input and $\mathbb{O}$ be
the output of $M$.  Then, ${\it ME}[U](M) =\log(|\mathbb{O}|)$.
\end{lemma}

\begin{reftheorem}{\ref{thm:meks}}
  $C_{\it ME}[U]$ is not a $k$-safety property for any $k>0$.
\end{reftheorem}
\begin{IEEEproof}
  For contradiction, suppose $C_{\it ME}[U]$ is a $k$-safety property.
  Let $M$ and $M'$ be programs having same input domain such that
  $(M,M')\not\in C_{\it ME}[U]$.  Then, it must be the case that there
  exist $T\subseteq\sembrack{M}$ and $T'\subseteq\sembrack{M'}$ such
  that $|T| \leq k$, $|T'| \leq k$, and $\forall M_c,M_c'.T \subseteq
  \sembrack{M_c} \wedge T' \subseteq \sembrack{M_c'} \Rightarrow (M_c,
  M_c') \not\in C_{\it ME}[U]$.
  
  Let 
\[
\begin{array}{l}
T=\aset{(h_1,o_1),(h_2,o_2),\dots,(h_i,o_i)}\\
T'=\aset{(h_1',o_1'),(h_2',o_2'),\dots,(h_j',o_j')}
\end{array}
\]
where $i,j\le k$.  Now, we construct new programs $\bar{M}$ and
$\bar{M'}$ as follows.
\[
\begin{array}{cc}
  \bar{M}(h_1)=o_1 &\bar{M'}(h_1')=o_1'\\
  \bar{M}(h_2)=o_2 &\bar{M'}(h_2')=o_2'\\
  \dots&\dots\\
  \bar{M}(h_i)=o_i &\bar{M'}(h_j')=o_j'\\
  \bar{M}(h_{i+1})=o &\bar{M'}(h_{j+1}')=o_{j+1}'\\
  \bar{M}(h_{i+2})=o &\bar{M'}(h_{j+2}')=o_{j+2}'\\
  \dots&\dots\\
  \bar{M}(h_{n})=o & \bar{M'}(h_{n}')=o_{n}'\\
\end{array}
\]
where 
\begin{itemize}
\item $o_{j+1}'$, $o_{j+2}'$, $\dots$, and $o_{n}'$ are distinct,
\item
  $\aset{o_1',o_2',\dots,o_j'}\cap\aset{o_{j+1}',\dots,o_{n}'}=\emptyset$,
\item $\aset{h_1,\dots,h_n} = \aset{h_1',\dots,h_n'}$, and
\item $n=2k$.
\end{itemize}
The number of outputs of the program $\bar{M'}$ is greater than or
equal to the number of the outputs of the program $\bar{M}$.  Hence,
by Lemma~\ref{lem:melog}, we have $(\bar{M},\bar{M'})\in C_{\it
ME}[U]$.  But, $T\subseteq\sembrack{\bar{M}}$ and
$T'\subseteq\sembrack{\bar{M'}}$.  This leads to a contradiction.
\end{IEEEproof}

\begin{definition}
\[
\begin{array}{l}
In(\mu,X,x)=|\aset{x'\in X\mid \mu(x')\ge\mu(x)}|
\end{array}
\]
\end{definition}
Intuitively, $In(\mu,X,x)$ is the order of $x$ defined in terms of
$\mu$.

\begin{lemma}
\[
\begin{array}{rl}
{\mathcal G}[\mu](X)&=\Sigma_{1\le i\le |X|}i\mu(x_i)\\
&=\Sigma_{x\in X}In(\mu,X,x)\mu(x)
\end{array}
\]
\end{lemma}
\begin{IEEEproof}
Trivial.
\end{IEEEproof}

\begin{lemma}
\label{lem:gelem}
  Let $\mu$ be a function such that $\mu : \mathbb{D}\rightarrow
  [0,1]$.  Let $P$ and $Q$ be sets such that $P\cup Q={\mathbb D}$ and
  $P\cap Q=\emptyset$.  Then, we have $\sum_{x\in
    \mathbb{D}}In(\mu,\mathbb{D},x)\mu(x)\ge\sum_{p\in
    P}In(\mu,P,p)\mu(p) + \sum_{q\in Q}In(\mu,Q,q)\mu(q)$.
\end{lemma}
\begin{IEEEproof}
Trivial.
\end{IEEEproof}

\begin{definition}
  Let $M$ be a function such that $M:\mathbb{A}\rightarrow
  \mathbb{B}$.  For any $o\in B$, we define $M^{-1}(o)$ to mean
\[
M^{-1}(o)=\aset{i\in \mathbb{A}\mid o=M(i)}
\]
\end{definition}

\begin{reftheorem}{\ref{thm:geks}}
 $C_{\it GE}[U]$ is not a k-safety property for any $k>0$
\end{reftheorem}
\begin{IEEEproof}
  For contradiction, suppose $C_{\it GE}[U]$ is a $k$-safety property.
  Let $M$ and $M'$ be programs having the same input domain such that
  $(M,M')\not\in C_{\it GE}[U]$.  Then, it must be the case that there
  exist $T\subseteq\sembrack{M}$ and $T'\subseteq\sembrack{M'}$ such
  that $|T| \leq k$, $|T'| \leq k$, and $\forall M_c,M_c'.T \subseteq
  \sembrack{M_c} \wedge T' \subseteq \sembrack{M_c'} \Rightarrow (M_c,
  M_c') \not\in C_{\it GE}[U]$.
  
Let 
\[
\begin{array}{l}
T=\aset{(h_1,o_1),(h_2,o_2),\dots,(h_i,o_i)}\\
T'=\aset{(h_1',o_1'),(h_2',o_2'),\dots,(h_j',o_j')}
\end{array}
\]
where $i,j\le k$.  Now, we construct new programs $\bar{M}$ and
$\bar{M'}$ as follows.
\[
\begin{array}{cc}
  \bar{M}(h_1)=o_1 &\bar{M'}(h_1')=o_1'\\
  \bar{M}(h_2)=o_2 &\bar{M'}(h_2')=o_2'\\
  \dots&\dots\\
  \bar{M}(h_i)=o_i &\bar{M'}(h_j')=o_j'\\
  \bar{M}(h_{i+1})=o &\bar{M'}(h_{j+1}')=o_{j+1}'\\
  \bar{M}(h_{i+2})=o &\bar{M'}(h_{j+2}')=o_{j+2}'\\
  \dots&\dots\\
  \bar{M}(h_{i+j})=o &\bar{M'}(h_{j+i}')=o_{j+i}'\\
  \bar{M}(h_{i+j+1})=o_r & \bar{M'}(h_{j+i+1}')=o_r'\\
  \dots&\dots\\
  \bar{M}(h_{n})=o_r & \bar{M'}(h_{n}')=o_r'\\
\end{array}
\]
where 
\begin{itemize}
\item $o\not=o_r$,
\item $\aset{o_1,o_2,\dots,o_i}\cap\aset{o,o_r}=\emptyset$,
\item $o_{j+1}'$, $o_{j+2}'$, $\dots$, $o_{j+i}'$, and $o_r'$ are
  distinct,
\item
  $\aset{o_1',o_2',\dots,o_j'}\cap\aset{o_{j+1}',\dots,o_{j+i}',o_r'}=\emptyset$,
\item $\aset{h_1,\dots,h_n} = \aset{h_1',\dots,h_n'}$, and
\item $n=2k$.
\end{itemize}
We compare the guessing-entropy-based quantitative information flow of
the two programs.
\[
\begin{array}{l}
  {\it GE}[U](\bar{M'})-{\it GE}[U](\bar{M})\\
  \quad=\frac{|\mathbb{H}|}{2}-\frac{1}{2|\mathbb{H}|}\sum_{o'\in M'(\mathbb{H})}|M'^{-1}(o')|^2\\
  \qquad-\frac{|\mathbb{H}|}{2}+\frac{1}{2|\mathbb{H}|}\sum_{o\in M(\mathbb{H})}|M^{-1}(o)|^2\\
  \quad=\frac{1}{2|\mathbb{H}|}\sum_{o\in M(\mathbb{H})}|M^{-1}(o)|^2\\
\qquad-\frac{1}{2|\mathbb{H}|}\sum_{o'\in M'(\mathbb{H})}|M'^{-1}(o')|^2\\
  \quad=\frac{1}{2|\mathbb{H}|}(\sum_{o_x\in\aset{o_1,\dots,o_i}}|M^{-1}(o_x)|^2\\
\qquad\qquad+|M^{-1}(o)|^2+|M^{-1}(o_r)|^2)\\
  \qquad-\frac{1}{2|\mathbb{H}|}(\sum_{o_x'\in\aset{o_1',\dots,o_j'}}|M'^{-1}(o_x')|^2\\
  \qquad\qquad+\sum_{o_y'\in\aset{o_{j+1}',\dots,o_{j+i}'}}|M'^{-1}(o_y')|^2\\
\qquad\qquad+|M'^{-1}(o_r')|^2)\\
\end{array}
\]
By lemma~\ref{lem:gelem}, we have
\[
\begin{array}{l}
  \sum_{o_x\in\aset{o_1,\dots,o_i}}|M^{-1}(o_x)|^2\\
\qquad\le\sum_{o_y'\in\aset{o_{j+1}',\dots,o_{j+i}'}}|M'^{-1}(o_x')|^2\\
\text{and}\\
  |M^{-1}(o)|^2\le\sum_{o_x'\in\aset{o_1',\dots,o_j'}}|M'^{-1}(o_x')|^2
\end{array}
\]
Trivially, we have
\[
|M'^{-1}(o_r')|^2=|M^{-1}(o_r)|^2
\]
As a result, we have
\[
{\it GE}[U](\bar{M'})-{\it GE}[U](\bar{M})\ge 0
\]
Recall that $\bar{M}$ and $\bar{M'}$ have the same counterexamples $T$
and $T'$, that is, $T\subseteq\sembrack{\bar{M}}$ and
$T'\subseteq\sembrack{\bar{M'}}$.  However, we have
$(\bar{M},\bar{M'})\in C_{\it GE}[U]$.  This leads to a contradiction.
\end{IEEEproof}

\begin{reftheorem}{\ref{thm:ccks}}
$C_{\it CC}$ is not a $k$-safety property for any $k > 0$.
\end{reftheorem}
\begin{IEEEproof}
Straightforward from Lemma~\ref{lem:ccme} and Theorem~\ref{thm:meks}.
\end{IEEEproof}

\begin{figure}
\[
\begin{array}{l}
T(\phi)=\\
\ \ {\sf if}\;\phi\\
\ \ \ \ \ \ {\sf then}\;O_f:={\sf true};\vect{O}:=\vect{H}\\
\ \ \ \ \ \ {\sf else}\;O_f:={\sf false};\vect{O}:=\vect{{\sf false}}
\end{array}
\]
where $O_f$ and $\vect O$ are distinct.
\caption{Boolean formula encoding by boolean program}
\label{boolenc}
\end{figure}

\begin{lemma}
\label{lem:a5}
Let $\vect H$ be distinct boolean variables, $\phi$ be a boolean
formula over $\vect H$, and $n$ be the number of satisfying
assignments for $\phi$.  If $n$ is less than $2^{|{\vect H}|}$, then
the number of the outputs of the boolean program $T(\phi)$ defined in
Figure~\ref{boolenc} is equal to $n+1$.
\end{lemma}
\begin{IEEEproof}
Trivial.
\end{IEEEproof}

\begin{lemma}
\label{lem:a6}
Let $\vect H$ be distinct variables and $\phi$ be a boolean formula
over $\vect H$.  Then, the number of assignments for $\phi$ can be
computed by executing an oracle that decides whether programs are in
$C_{\it ME}[U]$ at most $3*(|\vect H|+1)+2$ times.
\end{lemma}
\begin{IEEEproof}
First, we define a procedure that returns the number of solutions for
$\phi$.
  
Let $B(j)=\psi\wedge H'$ where $\psi$ is a formula over $\vect H$
having $j$ assignments and $H'$ is a boolean variable such that
$H'\not\in\aset{\vect H}$.  Note that by Lemma~\ref{lem:a2}, such
$\psi$ can be generated in linear time.

Then, we invoke the following procedure where $T$ is defined
in Figure~\ref{boolenc}.
\[
\begin{array}{l}
\ell=0;\\
r=2^{|\vect H|};\\
n=(\ell+r)/2;\\
{\sf while}\;\neg((T(\phi\wedge H'),T(B(n)))\in C_{\it ME[U]}\\
\qquad\qquad{\sf and}\;(T(B(n)),T(\phi\wedge H'))\in C_{\it ME}[U])\\
\qquad{\sf if}\;(T(\phi\wedge H'),T(B(n)))\in C_{\it ME}[U]\\
  \qquad\qquad{\sf then}\;\{\ell=n;n=(\ell+r)/2;\}\\
  \qquad\qquad{\sf else}\;\{r=n;n=(\ell+r)/2;\}\\
  {\sf return}\;n
\end{array}
\]

Note that when the procedure terminates, we have ${\it
ME}[U](T(B(n))={\it ME}[U](T(\phi\wedge H'))$, and so by
Lemma~\ref{lem:melog} and Lemma~\ref{lem:a5}, $n$ is the number of
satisfying assignments to $\phi$.

We show that the procedure iterates at most $|\vect{H}|+1$ times.  To
see this, note that every iteration in the procedure narrows the range
between $r$ and $\ell$ by one half.  Because $r - \ell$ is bounded by
$2^{|\vect H|}$, it follows that the procedure iterates at most
$|\vect{H}|+1$ times.  Hence, the oracle $C_{\it ME}[U]$ is accessed
$3*(|\vect{H}|+1)+2$ times, and this proves the lemma.
\end{IEEEproof}

\begin{reftheorem}{\ref{thm:mecomp}}
$\text{\#P}\subseteq \text{FP}^{C_{\it ME}[U]}$
\end{reftheorem}
\begin{IEEEproof}
  Straightforward by Lemma~\ref{lem:a6} and the fact that \#SAT, the
  problem of counting the number of solutions to a boolean formula, is
  \#P-complete.
\end{IEEEproof}

\begin{lemma}
\label{lem:gemono}
  Let $\vect H$ and $H'$ be distinct variables and $\phi$ and $\phi'$
  be boolean formulas over $\vect H$.  Let $M\equiv
  O:=\phi\wedge H'$ and $M'\equiv O:=\phi'\wedge H'$.  Then, we
  have $\#SAT(\phi)\le\#SAT(\phi')$ iff ${\it GE}[U](M)\le{\it
    GE}[U](M')$.
\end{lemma}
\begin{IEEEproof}
By the definition, 
\[
\begin{array}{rl}
  {\it GE}[U](M)&=\mathcal{G}(H)-\mathcal{G}(H|O)\\
  &=\frac{1}{2}(|\vect H|)+\frac{1}{2}-\sum_o\sum_{1\le i\le |\vect H|}i U(h_i,o)\\
  &=\frac{|\vect H|}{2}\\
&\quad -\frac{1}{2|\vect H|}(|M^{-1}({\sf true})|^2+|M^{-1}({\sf
    false})|^2)
\end{array}
\]
Therefore, 
\[
{\it GE}[U](M) \leq {\it GE}[U](M')
\]
iff 
\[
\begin{array}{l}
|M^{-1}({\sf true})|^2+|M^{-1}({\sf false})|^2\\
\quad\ge|M'^{-1}({\sf
  true})|^2+|M'^{-1}({\sf false})|^2
\end{array}
\]
But, trivially, the latter holds iff
\[
\#SAT(\phi)\le\#SAT(\phi')
\]
\end{IEEEproof}

\begin{lemma}
\label{lem:gecomp}
  Let $\vect H$ and $H'$ be distinct variables and $\phi$ be a boolean
  formula over $\vect H$.  Then, the number of assignments for $\phi$
  can be computed by executing an oracle that decides whether programs
  are in $C_{\it GE}[U]$ at most $3*(|\vect H|+1)+2$ times.
\end{lemma}
\begin{IEEEproof}
First, we define a procedure that returns the number of solutions for $\phi$.

Let $B(j)=\psi\wedge H'$ where $\psi$ is a formula over $\vect H$
having $j$ assignments and $H'$ is a boolean variable such that
$H'\not\in\aset{\vect H}$.  Note that by Lemma~\ref{lem:a2}, such
$\psi$ can be generated in linear time.
\[
\begin{array}{l}
\ell=0;\\
r=2^{|\vect H|};\\
n=(\ell+r)/2;\\
{\sf while}\;\neg(O:=\phi\wedge H',O:=B(n))\in C_{\it GE[U]}\\
\qquad\qquad{\sf and}\;(O:=B(n),O:=\phi\wedge H')\in C_{\it GE}[U])\\
\qquad{\sf if}\;(O:=\phi\wedge H',O:=B(n))\in C_{\it GE}[U]\\
  \qquad\qquad{\sf then}\;\{\ell=n;n=(\ell+r)/2;\}\\
  \qquad\qquad{\sf else}\;\{r=n;n=(\ell+r)/2;\}\\
{\sf return}\;n
\end{array}
\]

Note that when this procedure terminates, we have ${\it
GE}[U](O:=B(n))={\it GE}[U](O:=\phi\wedge H')$, and so by
Lemma~\ref{lem:gemono}, $n$ is the number of satisfying assignments to
$\phi$.

We show that the procedure iterates at most $|\vect{H}|+1$ times.  To
see this, every iteration in the procedure narrows the range between
$r$ and $\ell$ by one half.  Because $r - \ell$ is bounded by
$2^{|\vect H|}$, it follows that the procedure iterates at most
$|\vect{H}|+1$ times.  Hence, the oracle $C_{\it GE}[U]$ is accessed
$3*(|\vect{H}|+1)+2$ times, and this proves the lemma.
\end{IEEEproof}

\begin{reftheorem}{\ref{thm:gecomp}}
$\text{\#P}\subseteq \text{FP}^{C_{\it GE}[U]}$
\end{reftheorem}
\begin{IEEEproof}
  Straightforward by Lemma~\ref{lem:gecomp} and the fact that \#SAT,
  the problem of counting the number of solutions to a boolean
  formula, is \#P-complete.
\end{IEEEproof}

\begin{reftheorem}{\ref{thm:cccomp}}
$\text{\#P}\subseteq \text{FP}^{C_{\it CC}}$
\end{reftheorem}
\begin{IEEEproof}
Straightforward from Lemma~\ref{lem:ccme} and Theorem~\ref{thm:mecomp}.
\end{IEEEproof}

\begin{reftheorem}{\ref{thm:nicomp}}
Checking non-interference is coNP-complete for loop-free boolean programs.
\end{reftheorem}
\begin{IEEEproof}
  We write $\text{NI}$ for the decision problem of checking
non-interference of loop-free boolean programs.  We prove by reducing
$\text{NI}$ to and from UNSAT, which is coNP-complete.
\begin{itemize}
\item $\text{NI}\subseteq \text{UNSAT}$ 

We reduce via self composition~\cite{barthe:csfw04,darvas:spc05}.  Let
$M$ be a boolean program that we want to know if it is
non-interferent.  First, we make a copy of $M$, with each variable $x$
in $M$ replaced by a fresh (primed) variable $x'$.  Call this copy
$M'$.  Let $\phi = \wpre{M;M'}{\vect{O}=\vect{O}'}$, where
$\vect{O}=\vect{O}'$ is the boolean formula encoding the conjunction
of equalities $O_1 = O_1'$, $O_2 = O_2'$, \dots, $O_n = O_n'$, where
$O_1, \dots, O_n$ are the low security output variables of $M$.  Note
that $\phi$ can be obtained in time polynomial in the size of $M$.
Here, instead of the rules in Figure~\ref{wpsemantics}, we use the
optimized weakest precondition generation
technique~\cite{DBLP:conf/popl/FlanaganS01,DBLP:journals/ipl/Leino05}
that generates a formula quadratic in the size of $M;M'$.  Then, $M$
is non-interferent if and only if $\phi$ is valid, that is, if and
only if $\neg \phi$ is unsatisfiable.

\item $\text{UNSAT}\subseteq \text{NI}$ 

  Let $\phi$ be a formula that we want to know if it is unsatisfiable.
  We prove that the following programs is non-interferent iff $\phi$
  is unsatisfiable.  Here, all variables that appear in $\phi$ are
high security input variables and $H$ is a high security input variable
that is distinct from variables appearing in $\phi$, and $O$ is the
low security output variable.
\[
\begin{array}{l}
{\sf if}\;\phi\wedge H\;{\sf then}\;O:={\sf true}\;{\sf else}\;O:={\sf false}
\end{array}
\]
Trivially, if $\phi$ is unsatisfiable, then this program returns only
${\sf false}$, that is, this program is non-interferent.  If this
program is non-interferent, then this program returns only ${\sf
  true}$ for any input, or returns only ${\sf false}$ for any input.
However, this program can not return only ${\sf true}$, because if
$H={\sf false}$ then $\phi\wedge H={\sf false}$.  Therefore, this
program only returns ${\sf false}$, when this program is
non-interferent.  That means $\phi$ is unsatisfiable when the program
is non-interferent.
\end{itemize}
\end{IEEEproof}

\begin{definition}
  Let $M$ be a function such that $M:\mathbb{A}\rightarrow
  \mathbb{B}$.  Then, we define the image of $M$ on $\mathbb{X}\subseteq \mathbb{A}$, $M[\mathbb{X}]$, as follows.
\[
M[\mathbb{X}]=\aset{o\mid o=M(x)\wedge x\in \mathbb{X}}
\]
\end{definition}

\begin{lemma}
\label{lem:a10}
Let $\mathbb H$ be a set, and $M$ and $M'$ be functions whose domains
contain $\mathbb H$.  Suppose that we have
$M'(h_0,l)=M'(h_1,l)\Rightarrow M(h_0,l)=M(h_1,l)$, for all $h_0,h_1$
in $\mathbb H$.  Then, for all $h'\in\mathbb{H}$, we have
$\aset{h\mid M'(h,l)=M'(h',l)}\subseteq\aset{h\mid M(h,l)=M(h',l)}$.
\end{lemma}
\begin{IEEEproof}
Trivial.
\end{IEEEproof}
\texcomment{
\begin{lemma}
\label{lem:a11}
Let $\mathbb H$ be a set, and $M$ and $M'$ be functions whose domains
contain $\mathbb H$.  Let ${\mathbb H}_M=\aset{M^{-1}(M(h))\mid
  h\in\mathbb{H}}$ and ${\mathbb H}_{M'}=\aset{M'^{-1}(M(h))\mid
  h\in\mathbb{H}}$.  Suppose that we have $M'(h_0)=M'(h_1)\Rightarrow
M(h_0)=M(h_1)$, for all $h_0,h_1$ in $\mathbb H$.  Then, for all
$H,H'\in{\mathbb H}_{M'}$, if $H\not=H'$ then $H\cap H'=\emptyset$.
\end{lemma}
\begin{IEEEproof}
  Suppose $H\cap H'\not=\emptyset$.  Then, there exists $h$ such that
  $h\in H\cap H'$.  We have $H=M^{-1}(M(h))=H'$.  It leads a
  contradiction.  Therefore, $H\cap H'=\emptyset$.
\end{IEEEproof}
}

\begin{lemma}
\label{lem:a12}
Let $H$, $O$, $O'$, and $L$ be distinct random variables.  Let $M$ and
$M'$ be programs.  We have $(M,M')\in R$ iff for any distribution
$\mu$, $\mathcal{H}_\infty[\mu](H|O',L)\le
\mathcal{H}_\infty[\mu](H|O,L)$ where $O'=M'(H,L)$ and $O=M(H,L)$.
\end{lemma}
\begin{IEEEproof}
\begin{itemize}
\item ($\Rightarrow$)

Suppose $R(M,M')$.  We have
\[
\begin{array}{l}
  \mathcal{H}_\infty[\mu](H|O',L)\le \mathcal{H}_\infty[\mu](H|O,L)\\
  \qquad\textrm{ iff }\mathcal{V}[\mu](H|O,L)\le \mathcal{V}[\mu](H|O',L)
\end{array}
\]
by the definition of min entropy, and

\[
\begin{array}{l}
  \mathcal{V}[\mu](H|O,L)\\
\quad=\sum_{o\in{\mathbb O},\ell\in{\mathbb L}} \mu(o,\ell)\max_{h\in\mathbb H} \mu(h|o,\ell)\\
\quad  =\sum_{o\in{\mathbb O},\ell\in{\mathbb L}} \mu(o,\ell) \max_{h\in\mathbb H} \frac{\mu(h,o,\ell)}{\mu(o,\ell)}\\
\quad  =\sum_{o\in{\mathbb O},\ell\in{\mathbb L}} \max_{h\in\mathbb H} \mu(o,\ell)\frac{\mu(h,o,\ell)}{\mu(o,\ell)}\\
\quad  =\sum_{o\in{\mathbb O},\ell\in{\mathbb L}}\max_{h\in\mathbb H} \mu(h,o,\ell)\\
\quad  =\sum_{o\in{\mathbb O},\ell\in{\mathbb L}}\max_{h\in \aset{h'\mid o=M(h',\ell)}} \mu(h,\ell)\\
\end{array}
\]
where $\mathbb O=M[\aset{(h,\ell)\in\mathbb H\times\mathbb{L}\mid
  \mu(h,l)>0}]$, and ${\mathbb L}$ and ${\mathbb H}$ are sample spaces
  of low-security input and high-security input, respectively.
  Therefore, it suffices to show that
\[
\begin{array}{l}
  \mathcal{V}[\mu](H|O',L)-\mathcal{V}[\mu](H|O,L)\\
  \quad=\sum_{o'\in{\mathbb O'},\ell\in{\mathbb L}}\max_{h\in\aset{h'\mid o'=M'(h',\ell)}} \mu(h,\ell)\\
  \qquad-\sum_{o\in\mathbb O,\ell\in{\mathbb L}}\max_{h\in\aset{h'\mid o=M(h',\ell)}} \mu(h,\ell)\\
  \quad\ge 0
\end{array}
\]
where ${\mathbb O'}=M'[\aset{(h,\ell)\in\mathbb H\times\mathbb{L}\mid
  \mu(h,\ell)>0}]$.  

For any $o\in{\mathbb O}$ and $\ell \in \mathbb{L}$, there exists
$h_m$ such that $\mu(h_m,\ell)=\max_{h\in \aset{h'\mid
o=M(h',\ell)}}\mu(h,\ell)$.  Because $R(M,M')$, by
Lemma~\ref{lem:a10}, we have 
\[
\begin{array}{l}
\aset{h\mid M'(h,\ell)=M'(h_m,\ell)}\\
\hspace{3em}\subseteq\aset{h\mid
M(h,\ell)=M(h_m,\ell)}
\end{array}
\] 

Therefore, \[ \mu(h_m,\ell)=\max_{h\in\aset{h'\mid
o'=M'(h',\ell)}}\mu(h,\ell)
\] 
for some $o'\in {\mathbb O'}$.  Hence, each summand in
$\sum_{o\in{\mathbb O},\ell\in{\mathbb L}}\max_{h\in\aset{h'\mid
o=M(h',\ell)}} \mu(h,\ell)$ also appears in $\sum_{o'\in{\mathbb
O'},\ell\in{\mathbb L}}\max_{h\in \aset{h'\mid o'=M'(h',\ell)}}
\mu(h,\ell)$.  And, we have the above proposition.

\item($\Leftarrow$)

  We prove the contraposition.  Suppose $(M,M')\not\in R$.  Then,
  there exist $h_0,h_1,\ell,o_0,o_1$ such that
  $M'(h_0,\ell)=M'(h_1,\ell)$, $o_0=M(h_0,\ell)$, $o_1=M(h_1,\ell)$, and
  $o_0\not=o_1$.  Pick a probability distribution $\mu$ such that
  $\mu(h_0,\ell)=\mu(h_1,\ell)=\frac{1}{2}$.  Then, we have
\[
\begin{array}{l}
  \mathcal{V}[\mu](H|O',L)\\
\quad =\sum_{o'\in{\mathbb O'},\ell\in{\mathbb
      L}}\max_{h\in\aset{h'\mid o'=M(h',\ell)}} \mu(h,\ell)\\
  \quad= \frac{1}{2}
\end{array}
\]
and
\[
\begin{array}{l}
  \mathcal{V}[\mu](H|O,L)\\
\quad=\sum_{o\in{\mathbb O},\ell\in{\mathbb
      L}}\max_{h\in\aset{h'\mid o=M(h',\ell)}} \mu(h,\ell)\\
  \quad=  \frac{1}{2}+\frac{1}{2}\\
  \quad= 1
\end{array}
\]
Therefore, $\mathcal{H}_\infty[\mu](H|O',L)\not\le
\mathcal{H}_\infty[\mu](H|O,L)$.
\end{itemize}
\end{IEEEproof}

\begin{reftheorem}{\ref{thm:ReqME}}
  $R= \aset{(M_1,M_2) \mid \forall \mu.C_{\it ME}[\mu](M_1,M_2)}$
\end{reftheorem}
\begin{IEEEproof}
  Straightforward from Lemma~\ref{lem:a12} and the fact that
  $\mathcal{H}_\infty[\mu](H|L)-\mathcal{H}_\infty[\mu](H|O,L)\le
  \mathcal{H}_\infty[\mu](H|L)-\mathcal{H}_\infty[\mu](H|O',L)$ iff
  $\mathcal{H}_\infty[\mu](H|O,L)\ge\mathcal{H}_\infty[\mu](H|O',L)$.
\end{IEEEproof}

\begin{reftheorem}{\ref{thm:ReqGE}}
$R=\aset{(M_1,M_2)\mid\forall\mu.C_{\it GE}[\mu](M_1,M_2)}$
\end{reftheorem}
\begin{IEEEproof}
\begin{itemize}
\item $\subseteq$

\texcomment{
Suppose $(M,M') \in R$.  By the definition, 
\[
\begin{array}{l}
  {\it GE}[\mu](M)=\\
  \quad\sum_{\ell\in \mathbb{L}}\sum_{h\in\mathbb{H}}In(\lambda h'.\mu(h',\ell),{\mathbb H},h)\mu(h,\ell)\\
  \quad-\sum_{o\in\mathbb{O},\ell\in\mathbb{L}}\sum_{h\in\mathbb{H}}\\
  \qquad In(\lambda h'.\mu(h',o,\ell),\mathbb{H},h)\mu(h,o,\ell)
\end{array}
\]
and
\[
\begin{array}{l}
{\it GE}[\mu](M')=\\
\quad\sum_{\ell\in \mathbb{L}}\sum_{h\in\mathbb{H}}In(\lambda h'.\mu(h',\ell),{\mathbb H},h)\mu(h,\ell)\\
\quad-\sum_{o'\in\mathbb{O'},\ell\in\mathbb{L}}
\sum_{h\in\mathbb{H}}\\
\qquad In(\lambda h'.\mu(h',o',\ell),\mathbb{H},h)\mu(h,o',\ell)
\end{array}
\]
where ${\mathbb O}=M[\aset{(h,\ell)\in\mathbb{H}\times\mathbb{L}\mid
  \mu(h,\ell)>0}]$ and ${\mathbb
  O'}=M'[\aset{(h,\ell)\in\mathbb{H}\times\mathbb{L}\mid \mu(h,\ell)>0}]$.
}

Suppose $(M,M') \in R$.  By the definition, 
\[
\begin{array}{l}
  {\it GE}[\mu](M)=\\
\sum_{\ell\in \mathbb{L},h\in\mathbb{H}}In(\lambda h'.\mu(h',\ell),{\mathbb H},h)\mu(h,\ell)\\
-\sum_{o\in\mathbb{O},\ell\in\mathbb{L},h\in\mathbb{H}}
In(\lambda h'.\mu(h',o,\ell),\mathbb{H},h)\mu(h,o,\ell)
\end{array}
\]
and
\[
\begin{array}{l}
{\it GE}[\mu](M')=\\
\sum_{\ell\in \mathbb{L},h\in\mathbb{H}}In(\lambda h'.\mu(h',\ell),{\mathbb H},h)\mu(h,\ell)\\
-\sum_{o'\in\mathbb{O'},\ell\in\mathbb{L},h\in\mathbb{H}}
In(\lambda h'.\mu(h',o',\ell),\mathbb{H},h)\mu(h,o',\ell)
\end{array}
\]
where ${\mathbb O}=M[\aset{(h,\ell)\in\mathbb{H}\times\mathbb{L}\mid
  \mu(h,\ell)>0}]$ and ${\mathbb
  O'}=M'[\aset{(h,\ell)\in\mathbb{H}\times\mathbb{L}\mid \mu(h,\ell)>0}]$.

It suffices to show that
\[
\begin{array}{l}
\sum_{o'\in\mathbb{O'},\ell\in\mathbb{L},h\in\mathbb{H}}
In(\lambda h'.\mu(h',o',\ell),\mathbb{H},h)\mu(h,o',\ell)\\
\quad\le
\sum_{o\in\mathbb{O},\ell\in\mathbb{L},h\in\mathbb{H}}In(\lambda h'.\mu(h',o,\ell),\mathbb{H},h)\mu(h,o,\ell)
\end{array}
\]

\texcomment{
\[
\begin{array}{l}
\sum_{o'\in\mathbb{O'},l\in\mathbb{L}}\sum_{h\in\mathbb{H}}\\
\qquad In(\lambda h'.\mu(h',o',l),\mathbb{H},h)\mu(h,o',l)\\
\quad\le
\sum_{o\in\mathbb{O},l\in\mathbb{L}}\sum_{h\in\mathbb{H}}
\qquad In(\lambda h'.\mu(h',o,l),\mathbb{H},h)\mu(h,o,l)
\end{array}
\]
}

Let $o \in \mathbb{O}$ and $\ell \in \mathbb{L}$.  Let
$o=M(h_0,\ell)=\dots=M(h_x,\ell)$, and let
$o_0'=M'(h_0,\ell),\dots,o_x'=M'(h_x,\ell)$.  Because $R(M,M')$, for
any $h'$ such that $M'(h',\ell)\in\aset{o_0',\dots,o_x'}$, we have
$h'\in\aset{h_0,\dots,h_x}$.  Then, by Lemma~\ref{lem:gelem}, we have
\[
\begin{array}{l}
\sum_{h\in\mathbb{H_O}}In(\lambda
h'.\mu(h',o',\ell),\mathbb{H},h)\mu(h,o,\ell)\\
\quad\ge
\sum_{o'\in\mathbb{O'}_o,h\in\mathbb{H}_o}
In(\lambda h'.\mu(h',o',\ell),\mathbb{H},h)\mu(h,o',\ell)
\end{array}
\]
where 
\[
\begin{array}{rl}
\mathbb{O'}_o &= \aset{o_0',\dots,o_x'}\\
\mathbb{H}_o &=\aset{h_0,h_1,\dots,h_x}
\end{array}
\]

Now we prove each $\mathbb{O}_o$ constructed above are disjoint.  That
is, for $o_1$ and $o_2$ outputs of $M$ such that $o_1\not=o_2$,
$\mathbb{O}_{o_1}\cap \mathbb{O}_{o_2}=\emptyset$.  For a
contradiction, suppose $o'\in\mathbb{O}_{o_1}\cap \mathbb{O}_{o_2}$.
Then, there exist $h_1$ and $h_2$ such that $o_1=M(h_1,\ell)$,
$o'=M'(h_1,\ell)$, $o_2=M(h_2,\ell)$, and $o'=M'(h_2,\ell)$.  Since
$R(M,M')$, we have $o_1=o_2$, and it leads to a contradiction.  Hence,
we have for any $\ell\in\mathbb{L}$,
\[
\begin{array}{l}
\sum_{o'\in\mathbb{O'},h\in\mathbb{H}}
In(\lambda h'.\mu(h',o',\ell),\mathbb{H},h)\mu(h,o',\ell)\\
\quad\le
\sum_{o\in\mathbb{O},h\in\mathbb{H}}In(\lambda h'.\mu(h',o,\ell),\mathbb{H},h)\mu(h,o,\ell)
\end{array}
\]

\texcomment{
Each left-hand side term has corresponding right-hand side terms, and
corresponding right-hand side terms are disjoint.  Hence, we have for
any $\ell\in{\mathbb L}$
\[
\begin{array}{l}
\sum_{o'\in\mathbb{O'},h\in\mathbb{H}}In(\lambda h'.\mu(h',o',\ell),\mathbb{H},h)\mu(h,o',\ell)\\
\quad\le \sum_{o\in\mathbb{O},h\in\mathbb{H}}In(\lambda h'.\mu(h',o,\ell),\mathbb{H},h)\mu(h,o,\ell)
\end{array}
\]
}
Therefore, it follows that 
\[
\begin{array}{l}
\sum_{o'\in\mathbb{O'},\ell\in\mathbb{L},h\in\mathbb{H}}
In(\lambda h'.\mu(h',o',\ell),\mathbb{H},h)\mu(h,o',\ell)\\
\quad\le
\sum_{o\in\mathbb{O},\ell\in\mathbb{L},h\in\mathbb{H}}In(\lambda h'.\mu(h',o,\ell),\mathbb{H},h)\mu(h,o,\ell)
\end{array}
\]

\item $\supseteq$

  We prove the contraposition.  Suppose $(M,M')\not\in R$.  Then,
  there exist $h,h',\ell,o,o'$ such that
\begin{itemize}
\item $M(h,\ell)=o$, $M(h',\ell)=o'$, and $o\not=o'$
\item $M'(h,\ell)=M'(h',\ell)$
\end{itemize}
Then, we can pick $\mu$ such that $\mu(h,\ell)=\mu(h',\ell)=0.5$.  We have
\[
{\it GE}[\mu](M)=1.5-1=0.5
\]
and
\[
{\it GE}[\mu](M')=1.5-1.5=0
\]
Therefore, we have $(M,M')\not\in C_{\it GE}[\mu]$.
\end{itemize}
\end{IEEEproof}

\begin{reftheorem}{\ref{thm:RimpCC}}
$R \subseteq C_{\it CC}$
\end{reftheorem}
\begin{IEEEproof}
  Let $M$ and $M'$ be programs such that $(M,M')\in R$.  We prove
  $(M,M')\in C_{\it CC}$.  

By Theorem~\ref{thm:ReqSE}, we have
\[
\forall\mu. {\it SE}[\mu](M)\le{\it SE}[\mu](M')
\]
Now, there exists $\mu'$ such that
\[
{\it CC}(M) = {\it SE}[\mu'](M)
\]
Therefore,
\[
{\it SE}[\mu'](M)\le{\it SE}[\mu'](M')
\]
Trivially,
\[
{\it SE}[\mu'](M')\le {\it CC}(M')
\]
Therefore, we have the conclusion.
\end{IEEEproof}

\begin{reftheorem}{\ref{thm:rni}}
Let $M_2$ be a non-interferent program.  Then, $R(M_1, M_2)$ iff $M_1$
is also non-interferent and $M_1$ has the same input domain as $M_2$.
\end{reftheorem}
\begin{IEEEproof}
Straightforward from Theorems~\ref{thm:ReqNI} and \ref{thm:ReqSE}.
\end{IEEEproof}

\begin{reftheorem}{\ref{thm:rcomp}}
Restricted to loop-free boolean programs, $R$ is coNP-complete.
\end{reftheorem}
\begin{IEEEproof}
\begin{itemize}
\item $R\subseteq \text{coNP}$ 

  We prove by reducing $R$ to UNSAT, which is coNP-complete.  We
  reduce via self composition~\cite{barthe:csfw04,darvas:spc05}.  Let
  $M$ and $M'$ be boolean programs that we want to know if they are in
  $R$.  First, we make copies of $M$ and $M'$, with all variables in
  $M$ and $M'$ replaced by fresh (primed) variables.  Call these
  copies $M_c$ and $M_c'$.  Let $\phi =
  \wpre{M;M_c;M';M_c'}{\vect{O'}=\vect{O_c'} \Rightarrow
    \vect{O}=\vect{O_c}}$ where $\vect{O}$,$\vect{O_c}$,$\vect{O'}$,
  and $\vect{O_c'}$ are the low security outputs of $M$,$M_c$,$M'$,
  and $M_c'$, respectively.  Note that $\phi$ can be obtained in time
  polynomial in the size of $M$ and $M'$.  Here, like in
  Theorem~\ref{thm:nicomp}, we use the optimized weakest precondition
  generation
  technique~\cite{DBLP:conf/popl/FlanaganS01,DBLP:journals/ipl/Leino05}
  to generate a formula quadratic in the size of $M;M_c;M';M_c'$.
  Then, $(M,M')\in R$ if and only if $\phi$ is valid, that is, if and
  only if $\neg \phi$ is unsatisfiable.

\item $\text{coNP}\subseteq R$ 

  We prove by reducing NI to $R$, because NI is coNP-complete by
  Theorem~\ref{thm:nicomp}.  We can check the non-interference of $M$
  by solving $R(M,M')$ where $M'$ is non-interferent and have the
  same input domain as $M$ by Theorem~\ref{thm:rni}.  Note that such
  $M'$ can be constructed in polynomial time.  Therefore, we have
  $\text{coNP}\subseteq R$.
\end{itemize}
\end{IEEEproof}
\end{document}